\documentclass[lettersize,journal]{IEEEtran}
\usepackage{amsmath,amsfonts}
\usepackage{algorithmic}
\usepackage{array}
\usepackage[caption=false,font=normalsize,labelfont=sf,textfont=sf]{subfig}
\usepackage{textcomp}
\usepackage{stfloats}
\usepackage{url}
\usepackage{verbatim}
\usepackage{graphicx}
\def\BibTeX{{\rm B\kern-.05em{\sc i\kern-.025em b}\kern-.08em
    T\kern-.1667em\lower.7ex\hbox{E}\kern-.125emX}}
\usepackage{balance}
\usepackage{nicematrix}
\usepackage{multirow}
\usepackage{enumerate}
\usepackage[utf8]{inputenc}
\usepackage{graphicx}
\usepackage{float}
\usepackage{cite}
\usepackage{hyperref}
\usepackage{adjustbox}
\usepackage[skip=2pt]{caption} 
\usepackage{enumitem}
\usepackage{booktabs}
\usepackage{caption}
\usepackage{fancyhdr} 
\usepackage{xcolor}

\fancypagestyle{myfooter}{
  \fancyfoot[C]{\href{链接地址}{超链接内容}} 
}

\begin{document}
\title{Heterogeneous Graph Masked Contrastive Learning for Robust Recommendation}

\author{Lei Sang, Yu Wang, Yiwen Zhang*
\IEEEcompsocitemizethanks{\IEEEcompsocthanksitem Lei Sang, Yu Wang and Yiwen Zhang,  are with School of Computer Science and Technology, Anhui University 230601,
Hefei, Anhui, China.   E-mail: sanglei@ahu.edu.cn, wangyuahu@stu.ahu.edu.cn, zhangyiwen@ahu.edu.cn.
}

\thanks{*Corresponding author.}
}






\IEEEtitleabstractindextext{

\begin{abstract}
Heterogeneous graph neural networks (HGNNs) have demonstrated their superiority in exploiting auxiliary information for recommendation tasks. However, graphs constructed using meta-paths in HGNNs are usually too dense and contain a large number of noise edges. The propagation mechanism of HGNNs propagates even small amounts of noise in a graph to distant neighboring nodes, thereby affecting numerous node embeddings. To address this limitation, we introduce a novel model, named \textbf{M}asked \textbf{C}ontrastive \textbf{L}earning (MCL), to enhance recommendation robustness to noise. MCL employs a random masking strategy to augment the graph via meta-paths, reducing node sensitivity to specific neighbors and bolstering embedding robustness. Furthermore, MCL employs contrastive cross-view on a Heterogeneous Information Network (HIN) from two perspectives: one-hop neighbors and meta-path neighbors. This approach acquires embeddings capturing both local and high-order structures simultaneously for recommendation. Empirical evaluations on three real-world datasets confirm the superiority of our approach over existing recommendation methods.
\end{abstract}

}

\maketitle
\IEEEdisplaynontitleabstractindextext

\section{Introduction}


\IEEEPARstart {R} {ecommendation} systems \cite{cf, mf} find extensive application in diverse domains of everyday life, such as personalized shopping suggestions, tailored news delivery, and targeted social media content recommendations. Collaborative filtering (CF) \cite{mf}, a pivotal task in recommender systems, employs historical user-item bipartite graphs to derive vector representations for users and items. A significant approach within collaborative filtering is matrix factorization (MF) \cite{cf, ncf}. This technique represents users and items as latent feature vectors, enabling the anticipation of user-item interactions through linear combinations of these vectors. However, conventional recommendation methods primarily focus on predicting users' preferences based on their interactions with items; as a result, individual user attributes and item characteristics are overlooked, which frequently leads to a severe data sparsity problem.

\begin{figure}[t]
    \centering
    \includegraphics[width=\linewidth]{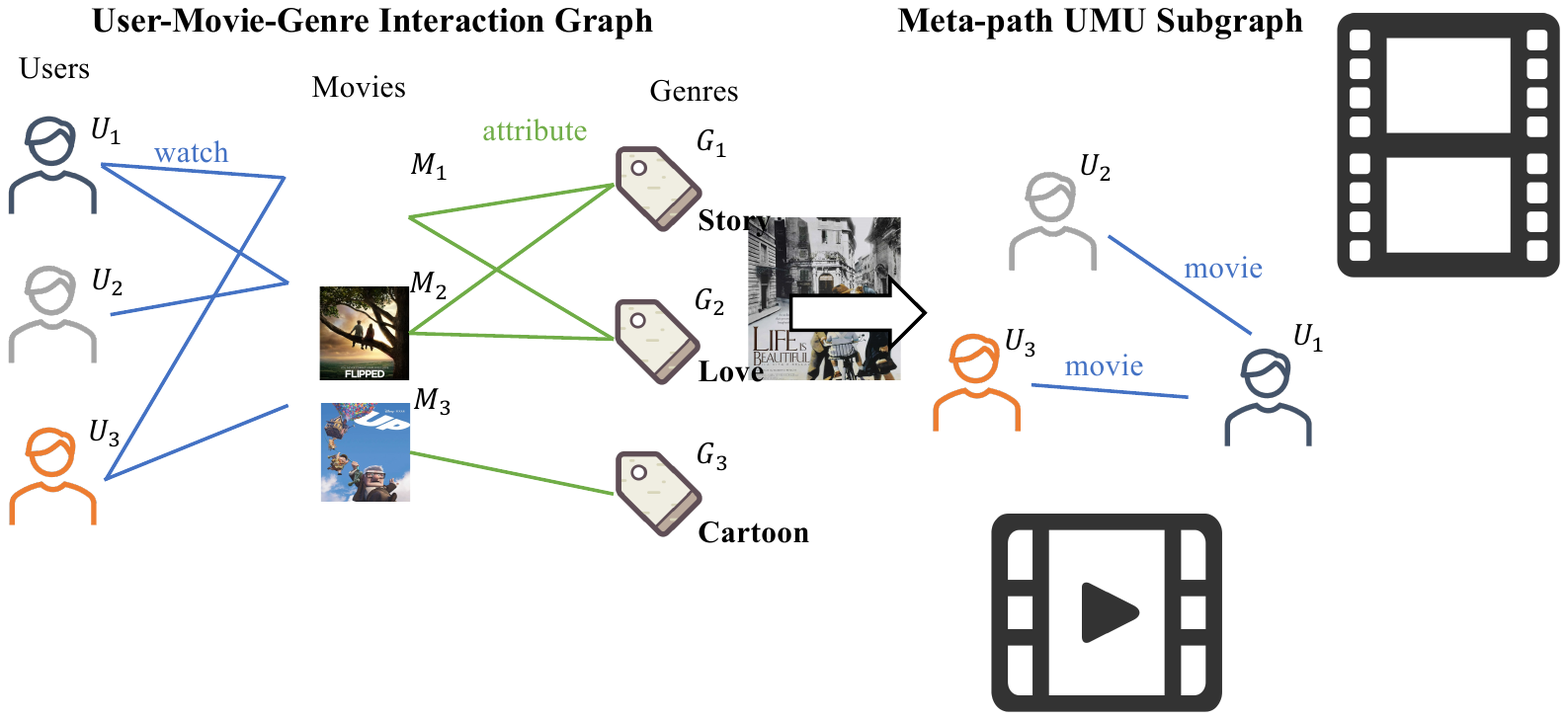}
    \caption{ An example of HIN. \textbf{Left}: There are three types of nodes: user, movie, and genre. There are two types of edges with different meanings: `U-M' (user to movie) and `M-G' (movie to genre). For the movie recommendation task, there are three different meta-paths: `U-M-U', `M-U-M', and `M-G-M'. \textbf{Right}: Users constitute a homogeneous graph through the defined meta-path `U-M-U'.}
    \label{fig1:intro}
\end{figure}

Heterogeneous Information Networks (HINs) \cite{hin, tahin} are a valuable tool for addressing data sparsity by incorporating additional user and item information. Unlike simple user-item graphs, HINs encompass diverse node and edge types, which expands their resemblance to real-world data. The present work treats such complex HINs as heterogeneous graphs, allowing for improved extraction of nuanced semantic and structural insights via meta-paths. For instance, Fig. \ref{fig1:intro} demonstrates that user U2 have never seen movie M1 before, while a connection can be established through the Genre node (via meta-path: U2-M2-G1-M1), motivating the incorporation of various meta-path types to mine hidden semantic interactions. The emerging field of heterogeneous graph recommendation, exemplified by models like MEIRec \cite{MEIrec} and CDPRec \cite{CDPrec}, frequently involves selecting higher-order homogeneous nodes using meta-path guidance. 
Such an intuition facilitates the inference of user preferences based on item/user similarity to generate effective recommendations. This concept is known as HIN-based recommendation.

The recent progress in Graph Neural Networks (GNNs) \cite{gat, gcn, random} provides a strong basis for HIN-based recommendation, especially through heterogeneous graph embedding models that automatically learn valuable features from raw input data. The main idea of GNN is to iteratively aggregate feature information from neighboring nodes and combine them with the representation of the current central node. Unlike traditional recommender systems, which only consider one-hop direct connectivity between a user and item, GNN-based recommender systems \cite{lightgcn,ngcf} can capture higher-order interactions in the user-item interaction graph and thereby obtain more optimal representations of users and items for recommendation. 
%
%
Currently, two prominent and effective techniques in the domain of HIN-based recommendation have emerged. 1) Extending GNNs to heterogeneous graphs results in Heterogeneous Graph Neural Networks (HGNNs) \cite{han, heco, rohe}, which address data sparsity via meta-paths and enhance semantic information by considering specific relationships \cite{han, heco}. While meta-paths generate more interactions, it is difficult to determine which of these are useful and which are noise. 2) In parallel, contrastive learning \cite{ssl2, ssl1} has been introduced to HGNN-based recommendation. Most of these methods \cite{hgcl, sgl} construct two similar views, perform cross-view alignment, and utilize InfoNCE loss \cite{infonce}, a self-supervised contrastive learning objective, to enhance the learning of representations. Furthermore, HGNN models \cite{hgcl, enhgcl}, built on contrastive learning principles, enhance representation distinctiveness, yielding discriminative user and item representations. However, existing HGNNs recommendation \cite{HERec, hgcl, smin} approaches still have certain the following challenges:

\textbf{CH1.} Despite the efficacy of HGNNs for recommendation, their performance is hindered by the escalating noise complexity. Meta-path-based interactions are typically supplement the sparse user-item interaction. However, the addition of more meta-paths does not always yield the improvement with more meta-paths. 
Unfortunately, HGNNs suffer from the phenomenon of perturbation enlargement \cite{rohe}, where introducing more edges can degrade their performance, as confirmed by our subsequent experiments.
Because HGNNs compute node embeddings by recursively aggregating information from neighborhoods, an iterative mechanism of this kind will have cascading effects: a small amount of noise in a graph will be propagated to neighboring nodes far away, affecting the embeddings of many other nodes.
Thus, one of the challenge with HGNN-based recommendation is effectively addressing noise complexity, which can impact performance as meta-paths are added, necessitating careful optimization strategies.

\textbf{CH2.} Furthermore, current contrastive learning (CL)-based methods \cite{sgl, hgcl, enhgcl, simgcl} encounter challenges when applied to heterogeneous graphs (HGs). Traditional CL-based recommendations \cite{sgl} construct multiple views using data augmentation while randomly sampling positive and negative samples based on user–item interactions. The recent HG-based recommendation method HGCL \cite{hgcl} also adopts this approach, leading to the neglect of rich semantic information and structural data. For example, heterogeneous graphs often contain meta-paths like `UMGMU' to connect two users. Discovering ways to use this information to get better positive samples and design effective contrastive learning strategies for such graphs is a valuable pursuit.

In this paper, to address the challenges outlined above, we introduce a novel model named \textbf{M}asked \textbf{C}ontrastive \textbf{L}earning (MCL). First, to counteract the substantial noise, MCL utilizes an embedding perturbation technique. This involves a random masking  strategy based on the Bernoulli distribution, retaining nodes with probability $p$ and masking them with probability ($1 - p$). 
HGNN's propagation operation for node representation learning is controlled by the mask, which can  reduce node sensitivity to specific neighbors and enhance embedding robustness.
Second, MCL designs cross-view contrastive learning on a Heterogeneous Information Network (HIN) from two perspectives: local view based on one-hop neighbors and global view based on 
meta-path neighbors. The local view obtains embeddings for users and items by aggregating one-hop heterogeneous neighbor nodes, while the global view captures embeddings of high-order homogeneous nodes using meta-paths. Finally, we propose a novel HG-based sampling strategy for contrastive learning, aiming to enhance the similarity between positive samples in cross-view contrast and to ensure that negative samples are far apart from each other.
%
Specifically, for the challenge and method, we give an model robustness analysis in the experimental section.
The contributions of this paper are as follows:
\begin{itemize}[leftmargin=*]
\item  We propose a novel model, named Masked Contrastive Learning (MCL), to enhance recommendation robustness against noise edge. MCL incorporates mask embedding and random propagation techniques to obtain enhanced embeddings. These embeddings are then aggregated using a two-level attention mechanism at both the node-level and semantic-level, effectively mitigating noise issues in recommendations.
\item We augment the supervised signal with two aligned views and employ a meta-path-based sampling approach, which can obtain high-quality positive and negative samples for recommendation.
\item We evaluate the top-K recommendation task using MCL on three real-world datasets. The experimental results demonstrate that our model achieves significant improvements compared to baseline methods when it comes to effectively extracting heterogeneous information and handling noise. We also achieve better results in the final robustness test.
\end{itemize}

\section{Preliminaries}
\noindent In this section, we  introduce key concepts essential for HIN-based recommendation, framed within the domain of heterogeneous information network, meta-path, meta-path-based subgraph, and heterogeneous information network embedding.

\textbf{Heterogeneous Information Network (HIN)}: A HIN is characterized by a graph structure $G=(V, E, I, R, \phi, \varphi)$, in which the collections of nodes and edges are symbolized by $V$ and $E$ respectively. For every node $v$ and edge $e$, there are associated type mapping functions $\phi: V \rightarrow I$ for nodes and $\varphi: E \rightarrow R$ for edges. Here, $I$ represents the node types and $R$ denotes the edge types, with the total number of types exceeding two, i.e., $|I|+|R|>2$.

\textbf{Network Schema}: The network schema, denoted as $T_{G}=(I, R)$, is used to demonstrate the high-level topology of an HIN, which describes the node types and their interaction relations. The network schema itself is a directed graph that is predicated on a collection of node types $I$, where the links between them are represented by the set of relational edge types $R$.

\textbf{Meta-path} \cite{meta-path}: In an $\operatorname{HIN} G=(V, E, I, R, \phi, \varphi)$, a meta-path $\rho$ is represented as $I_{1} \stackrel{R_{1}}{\longrightarrow} I_{2} \stackrel{R_{2}}{\longrightarrow} \ldots \stackrel{R_{l}}{\longrightarrow} I_{l+1}$, characterizing a composite connection between $v_{1}$ and $v_{l}$. Taking Fig. \ref{fig1:intro} as an instance, multiple meta-paths,  such as `Movie-User-Movie (MUM)'and `Movie-Genre-Movie (MGM)' can link entities within the HIN. Commonly, different meta-paths can reveal the different inter-dependency information of two movies. The path `MUM' indicates that two movies were watched by the same user, while `MGM' illustrates that two movies are of the same genre.

\textbf{Meta-path-based Subgraph}: Based on different meta-paths, we can extract homogeneous user-user and item-item interaction graphs from the original HIN, which are denoted as $\mathcal{G}{uu}$ and $\mathcal{G}{ii}$, respectively.

\textbf{Heterogeneous Information Network Embedding}: The embeddings $\mathbf{e}_{u}, \mathbf{e}_{i}, \mathbf{e}_{else}\in \mathbb{R}^{d}$  are initialized by Xavier \cite{xavier} uniform distribution, and $d$ is the hidden dimensionality. Each node is interconnected to form the embedding matrices $\mathbf{E}_{u}^{0} \in \mathbb{R}^{m \times d}$, $\mathbf{E}_{i}^{0} \in \mathbb{R}^{n \times d}$, and $\mathbf{E}_{else}^{0} \in \mathbb{R}^{l \times d}$. Here, $\mathit{m}$, $\mathit{n}$, and $\mathit{l}$ represent the number of users, items, and other types of nodes, respectively.

\section{Methodology}

\begin{figure*}[t]
    \centering
    \includegraphics[width=\linewidth]{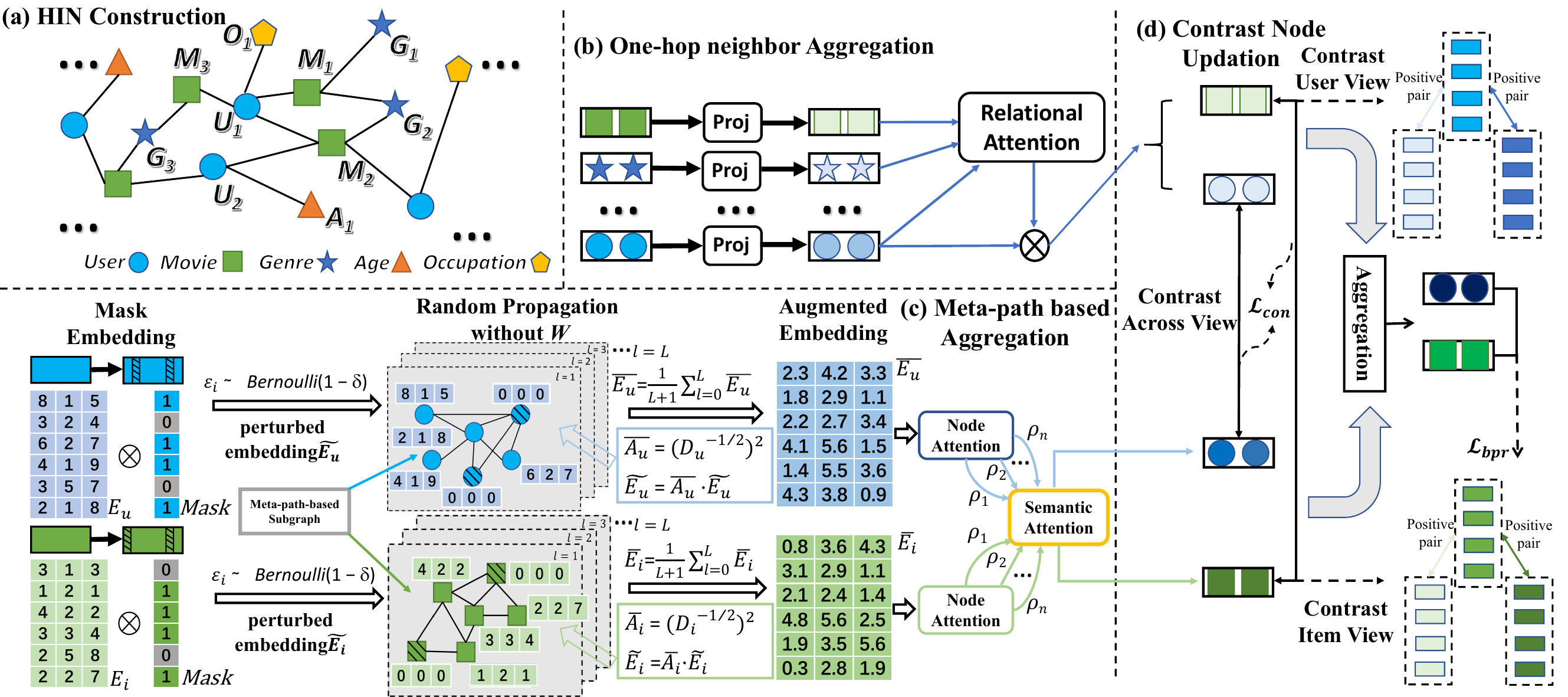}
    \caption{The overall framework of our proposed MCL. (a) Construction of heterogeneous information networks with multiple types of nodes and relations. (b) Mapping all types of nodes to the same space to aggregate information of one-hop neighbors. (c) Aggregation of nodes of the same type with random mask. (d) Using two-view contrastive learning to update node embeddings for final recommendation.}
    \label{fig3:framework}
\end{figure*}

\noindent In this section, we introduce the details of the proposed MCL, which consists of three modules designed to do the following: 1) project all types of nodes into the same node type space, aggregating information using only local one-hop neighbors; 2) obtain high-order neighbor information for aggregation based on global meta-path neighbors; 3) optimize the embedding learned from the two views by sampling and contrastive mechanisms. This results in tailored user and item embeddings for the recommendation task. The overall framework is presented in Fig. \ref{fig3:framework}.

\subsection{One-hop Neighbor based Aggregation}
\noindent Heterogeneous graphs offen contain diverse implications of various relations; for example, as shown in Fig. \ref{fig3:framework} (a), both $\text{U}_{1}$ and $\text{B}_{1}$ are neighbors of $\text{I}_{1}$, but these relations convey different meanings. We use relational attention mechanism to integrate the relationships among neighbor nodes within a one-hop distance. This process is illustrated in Fig. \ref{fig3:framework} (b), where $\boldsymbol{h}_{e}^{0}$ represents the embedding for each node in the concatenated set $\left\{{\mathbf{E}_{u}^{0}}, {\mathbf{E}_{i}^{0}}, {{E}_{else}^{0}}\right\}$. Due to the diversity of node types, we  project all nodes into the same space (denoted by $\boldsymbol{P}{\boldsymbol{r}}$ as the projection matrix) and then calculate the attention scores as follows:
$$
\alpha_{e w}=\frac{\exp \left(f_{r}\left(\boldsymbol{h}_{e}^{0}, \boldsymbol{P}_{r} \boldsymbol{h}_{w}^{0}\right)\right)}{\sum_{j \in \mathcal{N}_{1}(e)} \exp \left(f_{r}\left(\boldsymbol{h}_{e}^{0}, \boldsymbol{P}_{r} \boldsymbol{h}_{j}^{0}\right)\right)}, \quad \forall w \in \mathcal{N}_{1}(e),
$$
where $f_{r}(\cdot, \cdot)$ is the deep neural network that performs one-hop neighbor attention. Here, $\alpha_{ew}$ represents the influence level of node $e_w$, while $\mathcal{N}_{1}(e)$ denotes the set of one-hop neighboring nodes of $e$. Consequently, we aggregate the following information from $\mathcal{N}_{1}(e)$: 
\begin{equation}
\boldsymbol{h}_{e}^{1}=\sigma\left(\sum_{w \in \mathcal{N}_{1}(e)} \alpha_{e w} \boldsymbol{h}_{w}^{0}\right)
\end{equation}

\subsection{Meta-path based Aggregation}
\noindent We use an enhanced two-level attention network to extract two different types of structural information at the node level and the semantic level, respectively. 

\subsubsection{Node Level Attention with Random Mask}
MCL differs in important ways from some contrastive learning methods \cite{sgl, simgcl}: for example,  we do not use the topological information of the graph directly, nor do we change the original structure of the graph. First, without considering the graph structure, we randomly mask the entire embedding of a node $\mathbf{e}_{u}^{0}$ or $\mathbf{e}_{i}^{0}$ from $\mathbf{E}_{u}^{0}$ and $\mathbf{E}_{i}^{0}$ so that their initialized embeddings $\widetilde{\mathbf{E}_{u}^{0}}$ and $\widetilde{\mathbf{E}_{i}^{0}}$ are enhanced. The mask part uses Bernoulli distribution to generate a binary 0 or 1, i.e., $\widetilde{\mathbf{E}_{u}^{0}}=\frac{\epsilon_{i j}}{1-\delta} \mathbf{E}_{u}^{0}$, where $\epsilon_{i j}$ draws from Bernoulli $(1-\delta)$. In this process, we obtain the perturbed embedding matrices $\widetilde{\mathbf{E}_{u}^{0}}$ and $\widetilde{\mathbf{E}_{i}^{0}}$ by multiplying each node's embedding vector with its corresponding mask, i.e., $\widetilde{\mathbf{E}_{u}^{0}}=\mathbf{Mask} \cdot \mathbf{E}_{u}^{0}$, where $\mathbf{Mask}$ is a 0/1 matrix. In this way, we obtain the perturbed embedding $\widetilde{\mathbf{E}_{u}^{0}}$ and $\widetilde{\mathbf{E}_{i}^{0}}$ of the user and item. We then respectively define the degree matrices of $\mathcal{G}_{u u}$ and $\mathcal{G}_{i i}$ as $D_u$ and $D_i$. The random propagation equations on the graphs $\mathcal{G}_{u u}$ and $\mathcal{G}_{i i}$ and $D_u$ and $D_i$ obtained from the meta-paths are as follows:

\begin{equation}
\begin{aligned}
\left(\widetilde{E}^0\right)^{(L)} =  \bar{A} \cdot \widetilde{\mathbf{E}^{0}}, 
\bar{A} =  \left(D^{-\frac{1}{2}}\right)^2
\end{aligned}
\label{eq2}
\end{equation}

where we get a new $\widetilde{\mathbf{E}_{u}^{0}}$ and $\widetilde{\mathbf{E}_{i}^{0}}$ each time; these are used as the input for the next propagation. L iterations are performed before taking the average of the final augmented embeddings $\overline{\mathbf{E}_{u}^{0}}$ and $\overline{\mathbf{E}_{i}^{0}}$, Fig. \ref{fig3:framework} (c) illustrates the framework. We use the degrees of the nodes to represent their global importance and iterate through Eq. \ref{eq2} to obtain higher-order neighborhood information, reflecting the generalizability of our method. The homophily assumption suggests that adjacent nodes will tend to have similar features and labels \cite{homoli_assu}. Thus, the missing information of a node can be reconstructed using its neighbors, resulting in an approximate representation for the node in the corresponding augmentation:
\begin{equation}
\overline{\mathbf{E}^{0}} = \sum_{l=0}^L \frac{1}{L+1}\left(\widetilde{E}^0\right)^{(L)}
\end{equation}

Next, we pass the augmented embeddings $\overline{\mathbf{E}_{u}^{0}}$ and $\overline{\mathbf{E}_{i}^{0}}$ into the node-level attention network \cite{gat}. Here, $\boldsymbol{h}_{e}^{0}$ denotes every node on a certain kind of graph in $\mathcal{G}_{u u}$ or $\mathcal{G}_{i i}$. For example, $\mathcal{G}_{u u}$ uses the enhanced $\overline{\mathbf{E}_{u}^{0}}$.
$$
\beta_{e w}^{\rho}=\frac{\exp \left(f_{\rho}\left(\boldsymbol{h}_{e}^{0}, \boldsymbol{h}_{w}^{0}\right)\right)}{\sum_{j \in \mathcal{N}_{\rho}(e)} \exp \left(f_{\rho}\left(\boldsymbol{h}_{e}^{0}, \boldsymbol{h}_{j}^{0}\right)\right)}, \forall w \in \mathcal{N}_{\rho}(e),
$$
where the function $f_{\rho}(\cdot, \cdot)$ represents a neural network designed for node-level attention, and $\beta_{e w}$ represents the level of influence of node $e_{w}$. Subsequently, we employ an attention mechanism to effectively aggregate information from high-order neighbor nodes, following meta-paths specifically defined by experts:
\begin{equation}
\boldsymbol{h}_{e}^{\rho}=\sigma\left(\sum_{w \in \mathcal{N}_{\rho}(e)} \beta_{e w}^{\rho} \boldsymbol{h}_{w}^{0}\right)
\end{equation}
where $\sigma$ denotes the activation function.
\subsubsection{Semantic Level Attention}
After performing node attention aggregation on the given meta-path set $\left\{\rho_{1}, \rho_{2}, \ldots, \rho_{N}\right\}$, we obtain $N$ node-level embeddings for node $e$, which are represented as {$\left\{\boldsymbol{h}_{e}^{\rho_{1}}, {\boldsymbol{h}_{e}}^{\rho_{2}}, \ldots, {\boldsymbol{h}_{e}}^{\rho_{N}}\right\}$. Collectively, these node embeddings are denoted as $\left\{E_{\rho_{1}}, E_{\rho_{2}}, \ldots, E_{\rho_{N}}\right\}$. To enhance the accuracy of the learned node embeddings, we try to fuse multiple node embeddings. Taking $\left\{E_{\rho_{1}}, E_{\rho_{2}}, \ldots, E_{\rho_{N}}\right\}$ as input, as depicted in the right part of Fig. \ref{fig3:framework}} (c), we start by computing the significance of each meta-path $\rho_{j}$:
$$
w_{\rho_{j}}=\frac{1}{|\boldsymbol{V}|} \sum_{\boldsymbol{e} \in \mathcal{V}} \boldsymbol{q}^{T} \cdot \tanh \left(\boldsymbol{W} {\boldsymbol{h}_{e}}^{\rho_{j}}+b\right)
$$
where the weight for $\rho_{j},(j=1: N)$ is defined as follows:
$$
\gamma_{\rho_{j}}=\frac{\exp \left(w_{\rho_{j}}\right)}{\sum_{j=1}^{N} \exp \left(w_{\rho_{j}}\right)}.
$$
A larger value of $\gamma_{\rho_{j}}$ indicates a higher importance of the meta-path $\rho_{j}$. We then obtain the semantically weighted final node embeddings:
\begin{equation}
\boldsymbol{h}_{e}^{2}=\sum_{j=1}^{N} \gamma_{\rho_{j}} \cdot \boldsymbol{h}_{e}^{\rho_{j}}
\end{equation}

\subsection{Contrastive Views Learning}
\noindent In this section, we elaborate on Contrastive Views Learning from two perspectives. The first subsection describes the calculation of a similarity matrix between users and items under two different views, using contrastive loss and shared parameters. The second subsection introduces a multi-path sampling loss function used to enhance sampling accuracy.

\subsubsection{Similarity Matrix}
\noindent After obtaining the embeddings of users and items for both views, namely the one-hop neighbor view and meta-path neighbor view $\boldsymbol{h}_{e}^{1}$ and $\boldsymbol{h}_{e}^{2}$, we sample the interaction matrix of target nodes connected by many meta-paths to obtain the positive and negative samples that are used for the contrast. We map them into the space in which contrastive loss is calculated:
\begin{equation}
\begin{gathered}
\boldsymbol{e}^{1}{ }_{-} \text {proj} = \boldsymbol{W}^{(2)} \sigma\left( \boldsymbol{W}^{(1)} \boldsymbol{h}_{e}^{1}+b^{(1)}\right)+b^{(2)}, \\
\boldsymbol{e}^{2}{ }_{-} \text {proj} = \boldsymbol{W}^{(2)} \sigma\left( \boldsymbol{W}^{(1)} \boldsymbol{h}_{e}^{2}+b^{(1)}\right)+b^{(2)},
\end{gathered}
\end{equation}
where $\sigma$ is the ELU non-linear function. It is important to note that the parameters $W$ and $b$ are shared across the two views' embeddings. 

Next, we use $ \boldsymbol{e}_{u}^{1}{ }_{-} \text {proj}$ to refer to users under the one-hop neighbor view and $ \boldsymbol{e}_{u}^{2}{ }_{-} \text {proj}$ to refer to users under the meta-path neighbor view:
\begin{equation}
\begin{gathered}
\boldsymbol {sim}_{u} = \operatorname{softmax}(\exp (\operatorname{sim}(\boldsymbol{e}_{u}^{1}{ }_{-} \text {proj}, \boldsymbol{e}_{u}^{2}{ }_{-} \text {proj}) / \tau))
\end{gathered}
\end{equation}
where $\operatorname{sim}(u, v)$ denotes the cosine similarity between two vectors $\mathrm{u}$ and $\mathrm{v}$, while $\tau$ denotes a temperature parameter. Smaller temperature coefficients enable the separation of difficult samples that are similar to the present sample, resulting in a more uniform representation. Finally, according to the complementary nature of the two views, the similarity matrix is ${sim}_{u}^{1}$ from the one-hop neighbor's view; similarly, ${sim}_{u}^{2}$ is the transpose of ${sim}_{u}^{1}$ from the meta-path-based view. 

\subsubsection{Multi-path Sampling Loss}
\noindent In light of heterogeneous graphs' semantic richness and interaction complexity, and inspired by HeCo \cite{heco}, we  devised a suitable loss function for heterogeneous recommendation. First, we define user and item interaction matrices as zero matrices of \( m \times m \) and \( n \times n \), respectively. Next, we assign a value of 1 to the positive samples in the interaction matrix; in contrast, the remaining samples (with a value of 0 in the interaction matrix) are considered negative samples. This approach allows us to obtain two sampling matrices, ${Pos}_{u}$ and ${Pos}_{i}$, which contain positive samples for user-user and item-item interactions based on all meta-paths that interact with each other. For example, in Fig. \ref{fig1:intro}, the gray user is able to connect to the orange user through meta-path like `U-M-G-M-U', but they are not connected through `U-M-U'. Accordingly, these two users do not satisfy the definition and cannot be considered a positive sample pair. This method is employed to improve sampling accuracy and mitigate the impact of noise.
\begin{equation}
\boldsymbol {score}_{u} =(\text{Pos}_{u} \cdot \boldsymbol {sim}_{u})
\end{equation}

By multiplying the two matrices, we obtain two perspectives of the score matrix. A higher score indicates greater proximity to the positive sample. This process differs from the general recommendation contrastive loss function \cite{infonce}, as we utilize multiple pairs of positive samples that interact on all meta-paths. This approach enhances the quality of positive samples and implicitly utilizes negative samples.
We design the corresponding loss function as follows:
\begin{equation}
\mathcal{L}_{u}=-(\lambda_{1} \log \boldsymbol {score}_{u}^{1}+(1-\lambda_{1}) \log
\boldsymbol {score}_{u}^{2})
\end{equation}
where the coefficient $\lambda_{1}$ is introduced to balance the impact of the two views. By employing back-propagation, we can optimize the proposed model and acquire the node embeddings through the learning process. Similarly, the item also incur the loss $\mathcal{L}_{i}$ using the same strategy for both views.

\subsection{Optimization Objectives of MCL}
To optimize our MCL for the recommendation task, we use the Bayesian Personalized Ranking (BPR) \cite{bpr} pairwise loss function. The criterion for determining whether training samples are positive or negative is based on the presence or absence of user interaction. For example, a positive item $i$ is a product that has been purchased by the user, and a negative item $j$ is a product that has not been purchased. BPR \cite{bpr} assumes that observed interactions provide better reflections of user preferences compared to unobserved interactions and thus should be assigned higher predicted values. The objective function is defined as follows:
\begin{equation}
\mathcal{L}_{b p r}=\sum_{(u, i, j) \in O}-\ln \sigma\left(\hat{y}_{u i}-\hat{y}_{u j}\right)+\lambda_{2}\|\Theta\|_{2}^{2},
\end{equation}
where $\sigma$ denotes the sigmoid function. The hyperparameter $\lambda_{2}$ is used to control the weight of the $L_2$ regularization. The complete training loss -- which incorporates the losses, $\mathcal{L}_{u}$ and $\mathcal{L}_{i}$, obtained from supervised comparisons of user and item pairs in two separate views, into the joint training process with the BPR \cite{bpr} loss function -- can be expressed as follows:
\begin{equation}
\mathcal{L}=\mathcal{L}_{b p r}+\beta * \left(\mathcal{L}_{u} + \mathcal{L}_{i}\right)
\end{equation}
\section{Experiments}
\noindent In this section, we first evaluate our MCL and baseline methods and present theories and explanations for the experimental results. We then test the robustness of the model and the role of key modules and analyze the reasons for the observed experimental phenomena. Finally, we investigate the effect of hyperparameters in the model and conduct visualizations  of user and item representations. The following are the three main questions that we answer:

\begin{itemize}[leftmargin=*]
\item\textbf{RQ1}: How does MCL perform w.r.t. top -$K$ recommendation compared with existing methods?

\item\textbf{RQ2}: How does the robust performance of MCL in the face of noisy interactions and redundant meta-paths?


\item\textbf{RQ3}: Are the key components in our MCL delivering the expected recommendation performance gains?
\end{itemize}

\subsection{Experimental Setup}
\subsubsection{Datasets}
\noindent In our experiments, we use three real-world datasets that are publicly accessible and vary in terms of domain, size, and sparsity. Due to the limitations of heterogeneous graphs, we have had to select the current dataset as the main subject for experimentation, even as we incorporate more meta-paths to validate robustness. Additionally, the processed MovieLens dataset only exists in a 100k version. The size of the three datasets -- Movielens \footnote{\url{https://grouplens.org/datasets/movielens/}}, Amazon \cite{HERec}, and Yelp \footnote{\url{https://www.yelp.com/dataset/}} -- is increased sequentially. We present the details of each dataset in Table \ref{table1:datasets statistics}.

For each dataset, $80\%$ of the historical interactions per user are randomly selected as the training set, while the remaining interactions constitute the test set. From the training set, $10\%$ of the interactions are randomly sampled for the validation set (for hyperparameter tuning). Positive samples are defined as observed user-item interactions, while negative samples are generated through a sampling strategy that involves pairing each positive sample with an unobserved item previously not consumed by the user.

\begin{table}
\captionsetup{justification=centering}
\caption{Statistics of datasets used in experiments}
\begin{adjustbox}{width=0.48\textwidth}
\begin{tabular}{lllll}
\midrule Dataset & User \# & Item \# & Interaction \# & Sparsity \\
\midrule Movielens & 943 & 1,682 & 100,000 & $93.6953 \%$ \\
Amazon & 6,170 & 2,753 & 195,791 & $98.8473 \%$ \\
Yelp & 16,239 & 14,284 & 198,397 & $99.9145 \%$ \\
\midrule
\end{tabular}
\end{adjustbox}
\label{table1:datasets statistics}
\end{table}

\subsubsection{Baselines}
\noindent To evaluate the performance of our MCL, we compare it with several state-of-the-art models, which serve as baselines. 

\begin{itemize}[leftmargin=*]
    \item \textbf{Heterogeneous Graph Neural Network Models:}  HAN \cite{han}, HGT \cite{hgt}, HeCo \cite{heco} and RoHe \cite{rohe}.
    \item \textbf{Heterogeneous Graph Recommendation Models:} HERec \cite{HERec}, HAN \cite{han}, KGAT \cite{kgat}, NGCF+ \cite{ngcf}, SMIN \cite{smin} and HGCL \cite{hgcl}.
\end{itemize}

\begin{table*}[t]
    \captionsetup{justification=centering}
    \caption[Performance Comparison of Different Methods Using the Original Datasets]{%
    \textbf{Performance Comparison of Different Methods Using the Original Datasets}\\
    \small The best results are shown in bold and the second-best results are underlined. `Improve' denotes the relative improvement of MCL over the second-best results. Our MCL outperforms all existing baselines.}
    \begin{adjustbox}{width=\textwidth}
    \begin{tabular}{l|cccccccccccc}
    \toprule[1pt]
    \midrule
    \multirow{1}*{ Dataset } & \multicolumn{4}{c|}{ Movielens } & \multicolumn{4}{c|}{ Amazon } & \multicolumn{4}{c|}{ Yelp } \\ \midrule
    \multirow{2}*{ Metrics } & \multicolumn{2}{c|}{ @10 } & \multicolumn{2}{c|}{ @20 } & \multicolumn{2}{c|}{ @10 } & \multicolumn{2}{c|}{ @20 } & \multicolumn{2}{c|}{ @10 } & \multicolumn{2}{c|}{ @20 } \\ \cmidrule{2-13}
    ~ & Recall & NDCG & Recall & \multicolumn{1}{c|}{ NDCG } & Recall & NDCG & Recall & \multicolumn{1}{c|}{ NDCG } & Recall & NDCG & Recall & \multicolumn{1}{c|}{ NDCG } \\
    \midrule (2018)HERec & 0.2028 & 0.3692 & 0.3175 & 0.3747 & 0.0814 & 0.0766 & 0.1355 & 0.0983 & 0.0502 & 0.0355 & 0.0775 & 0.0451 \\
    \midrule (2019)KGAT & 0.2039 & 0.3711 & 0.3013 & 0.3586 & 0.0802 & 0.0750 & 0.1281 & 0.0918 & 0.0519 & 0.0361 & 0.0829 & 0.0499 \\
    \midrule (2019)NGCF+ & 0.1856 & 0.3330 & 0.2841 & 0.3371 & 0.0758 & 0.0661 & 0.1203 & 0.0802 & 0.0489 & 0.0372 & 0.0777 & 0.0467 \\
    \midrule (2019)HAN & 0.1773 & 0.3282 & 0.2876 & 0.3414 & 0.0748 & 0.0681 & 0.1206 & 0.0816 & 0.0456 & 0.0362 & 0.0773 & 0.0458 \\
    \midrule (2020)HGT & 0.2055 & 0.3682 & 0.3169 & 0.3744 & 0.0686 & 0.0623 & 0.1137 & 0.0781 & 0.0504 & 0.0362 & 0.0809 & 0.0461 \\
    \midrule (2021)HeCo & 0.2072 & 0.3787 & 0.3103 & 0.3726 & 0.0630 & 0.0584 & 0.1065 & 0.0726 & 0.0534 & 0.0400 & 0.0808 & 0.0460 \\
    \midrule (2021)SMIN & 0.1907 & 0.3437 & 0.3040 & 0.3628 & 0.0724 & 0.0634 & 0.1218 & 0.0824 & 0.0520 & 0.0388 & 0.0814 & 0.0475 \\
    \midrule (2022)RoHe & 0.1926 & 0.3468 & 0.3046 & 0.3582 & 0.0784 & 0.0720 & 0.1348 & 0.0918 & 0.0502 & 0.0367 & 0.0799 & 0.0465 \\
    \midrule (2023)HGCL & $\underline{0.2085}$ & $\underline{0.3790}$ & $\underline{0.3353}$ & $\underline{0.3921}$ & $\underline{0.0890}$ & $\underline{0.0813}$ & $\underline{0.1424}$ & $\underline{0.0991}$ & $\underline{0.0543}$ & $\underline{0.0408}$ & $\underline{0.0913}$ & $\underline{0.0538}$ \\
    \midrule (ours)MCL & $\mathbf{0.2193}$ & $\mathbf{0.3966}$ & $\mathbf{0.3415}$ & $\mathbf{0.4110}$ & $\mathbf{0.0920}$ & $\mathbf{0.0829}$ & $\mathbf{0.1441}$ & $\mathbf{0.1012}$ & $\mathbf{0.0596}$ & $\mathbf{0.0445}$ & $\mathbf{0.0961}$ & $\mathbf{0.0563}$ \\
    \midrule Improve & 5.18\% & 4.64\% & 1.85\% & 4.82\% & 3.37\% & 1.97\% & 1.19\% & 2.12\% & 9.76\% & 9.07\% & 5.26\% & 4.65\% \\
    \midrule
    \bottomrule[1pt]
    \end{tabular}
    \end{adjustbox}
    \label{table2:clean results}
\end{table*}

\subsubsection{Implementations}
\noindent For all baseline models, we prioritize the use of code provided by the original authors (if it exists pytorch version) or implemented through OpenHGNN \cite{openhgnn}, which is a unified open-source framework for developing and replicating heterogeneous graph algorithms. Some of these models were initially employed for heterogeneous graph node classification tasks (e.g., HAN, HeCo, HGT). In this context, we uniformly substitute their loss functions with the Bayesian Personalized Ranking (BPR) \cite{bpr} loss, while keeping the overall framework unchanged. We respectively set the sizes of the embedding and training batch to 128 and 2048 to ensure fair comparison. We use the Adam optimizer as a model-generic optimizer and initialize each model's parameters with the Xavier distribution.
Each model calculates preference scores for all items related to each user in the test set, excluding the positive samples from the training set. The top-$K$ recommendation approach is employed, which involves recommending $K$ items to the test users for evaluation. Evaluation metrics such as recall@$K$ and NDCG@$K$ are used, with $K$ set to 20.

\subsection{Performance Comparison (RQ1)}
\noindent Using Table \ref{table2:clean results}, we intend to show the performance of our MCL from two perspectives: recommendation performance and contrastive learning methods. In the first segment, we analyze why each method in the table reflects the performance. Subsequently, we analyze how the contrastive learning method of MCL differs from the general contrastive learning method.

\subsubsection{Performance Analysis}
\begin{itemize}[leftmargin=*]
\item First, we can see from Table \ref{table2:clean results} that MCL demonstrates superior recommendation performance across all clean datasets, exhibiting its most remarkable enhancement on the Yelp dataset (i.e., a relative improvement of up to 9.76\%). SMIN \cite{smin}, HeCo \cite{heco}, and HGCL \cite{hgcl} all employ self-supervised contrastive learning as an auxiliary task, leading to a substantial performance enhancement. These three methods construct views of contrastive learning in different ways: SMIN generates loss by contrasting the embedding obtained from the meta-path-based subgraphs and user-item bipartite graph learning; HGCL similarly combines meta network augmentation with a symmetric structure to formulate contrastive learning; finally, HeCo learns node embeddings through two views (network pattern view and meta-path view) of a HIN (Heterogeneous Information Network), effectively capturing both local and higher-order structures. HAN has become a widely employed heterogeneous graph learning framework in recent years, and RoHe \cite{rohe} shares the same framework, which employs an attention mechanism to defend against attacks. HERec \cite{HERec} and MCRec \cite{MCRec}, as earlier work, also exhibit promising recommendation performance after using BPR loss, indicating the continued relevance and potential of fusing meta-paths.

\item These different strategies also achieve different results on different datasets. For example, RoHe demonstrates relatively impressive performance on the Amazon and Yelp datasets. However, its performance drops declines the Movielens-100k dataset. One possible reason for this drop is that RoHe seems more suitable for sparser datasets. Because user-item interaction is less informative, both are significantly improved compared to HAN. HGCL, the most recent work, utilizes user-item interaction information directly, meaning that it always achieves relatively good results.
\end{itemize}

\begin{table}[t]
\caption{Performance comparison of contrastive learning methods}
\begin{adjustbox}{width=0.48\textwidth}
\begin{tabular}{ccccccc}
\toprule[1pt]
\midrule Data & \multicolumn{2}{|c|}{ Movielens } & \multicolumn{2}{c|}{ Amazon } & \multicolumn{2}{c}{ Yelp } \\
\midrule Metric & Recall & NDCG & Recall & NDCG & Recall & NDCG \\
\midrule MCL-InfoNCE & 0.3361 & 0.3998 & 0.1382 & 0.0956 & 0.0944 & 0.0549 \\
\midrule MCL & $\mathbf{0.3415}$ & $\mathbf{0.4109}$ & $\mathbf{0.1441}$ & $\mathbf{0.1012}$ & $\mathbf{0.0961}$ & $\mathbf{0.0563}$ \\
\midrule
\bottomrule[1pt]
\end{tabular}
\end{adjustbox}
\label{table4:contrast}
\vspace{-0.8em}
\end{table}

\subsubsection{Contrastive Learning of MCL}
\noindent To demonstrate the superiority of MCL's contrastive learning method in the field of heterogeneous graph recommendation, we substituted our contrastive learning approach with HGCL's InfoNCE loss and sampling method for contrastive learning. The outcomes are presented in Table \ref{table4:contrast}. As is clear from the table, we have effectively enhanced the selection of positive samples through multiple meta-paths for this specific definition of heterogeneous graph meta-paths. To illustrate this concept with a practical example related to social recommendation, consider the Movielens dataset: when users within the same age group and with the same occupation watch movies of identical genres, we posit that their movie-watching preferences align. Consequently, we leverage this premise to enhance their similarity from a user-embedding perspective. A similar principle holds when viewed from the movie perspective.

\begin{figure*}[t]
    \centering
    \begin{minipage}[b]{0.33\linewidth}
        \centering
        \includegraphics[width=\linewidth]{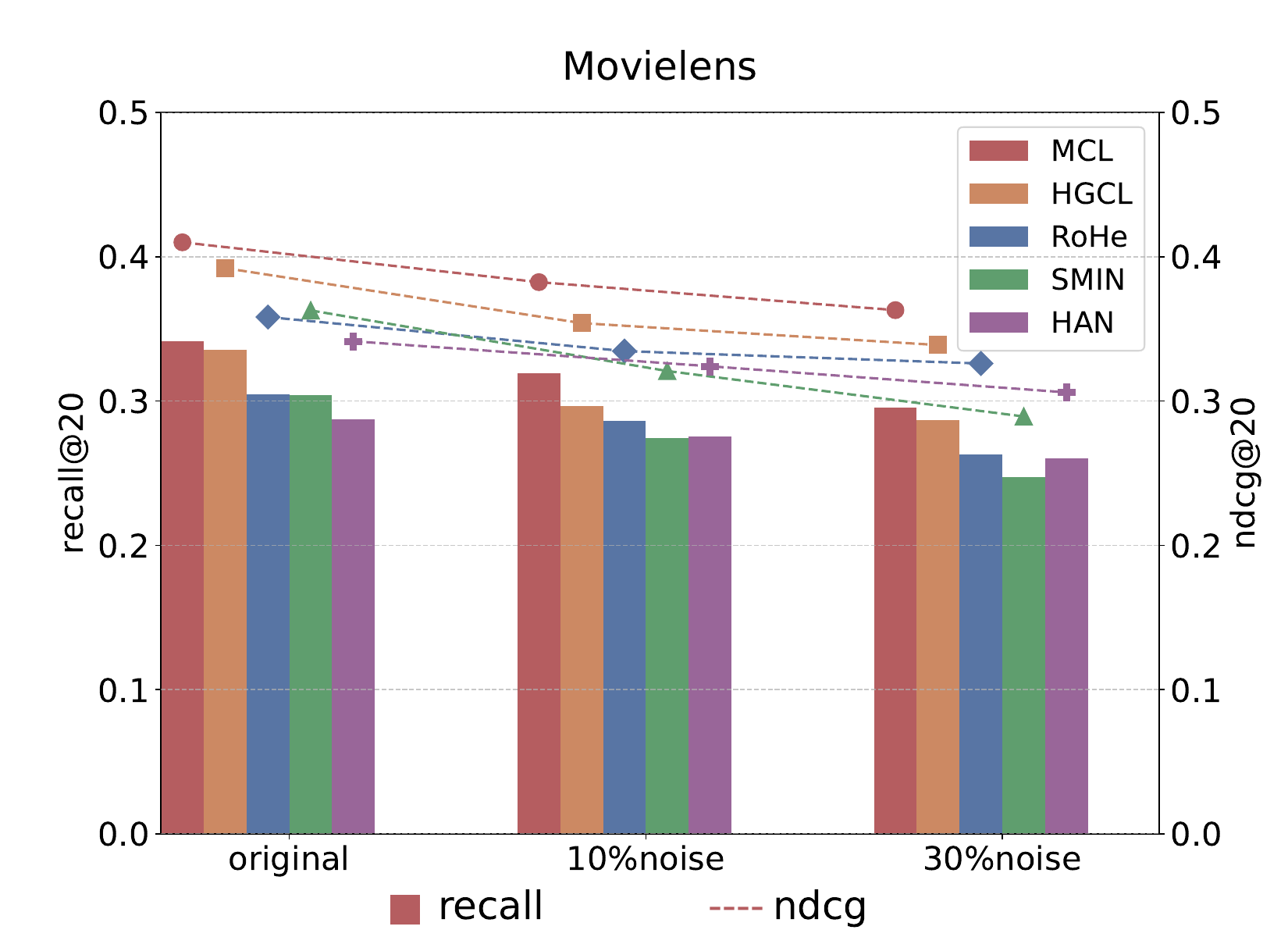}
    \end{minipage}\hfill
    \begin{minipage}[b]{0.33\linewidth}
        \centering
        \includegraphics[width=\linewidth]{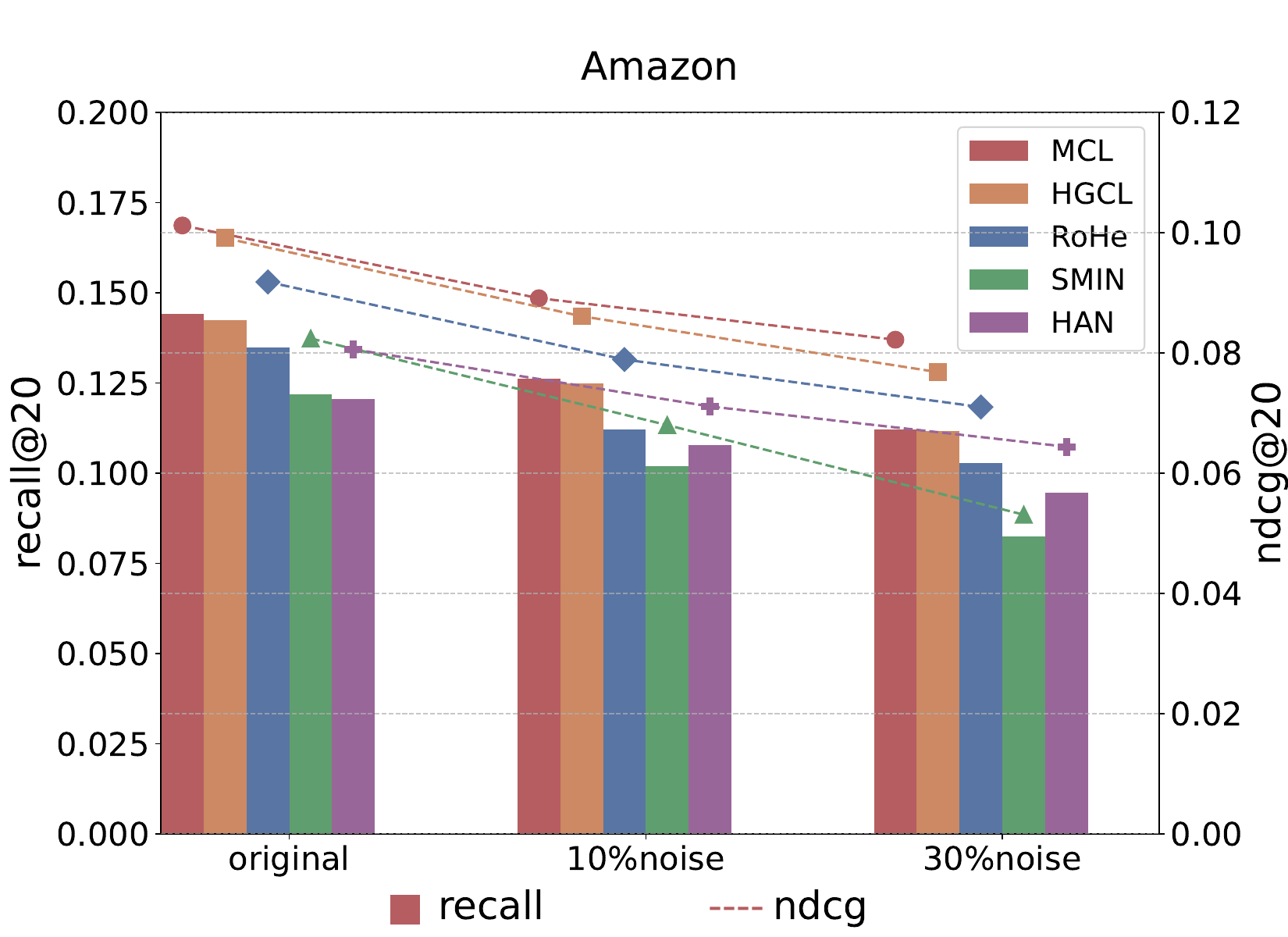}
    \end{minipage}
    \begin{minipage}[b]{0.33\linewidth}
        \centering
        \includegraphics[width=\linewidth]{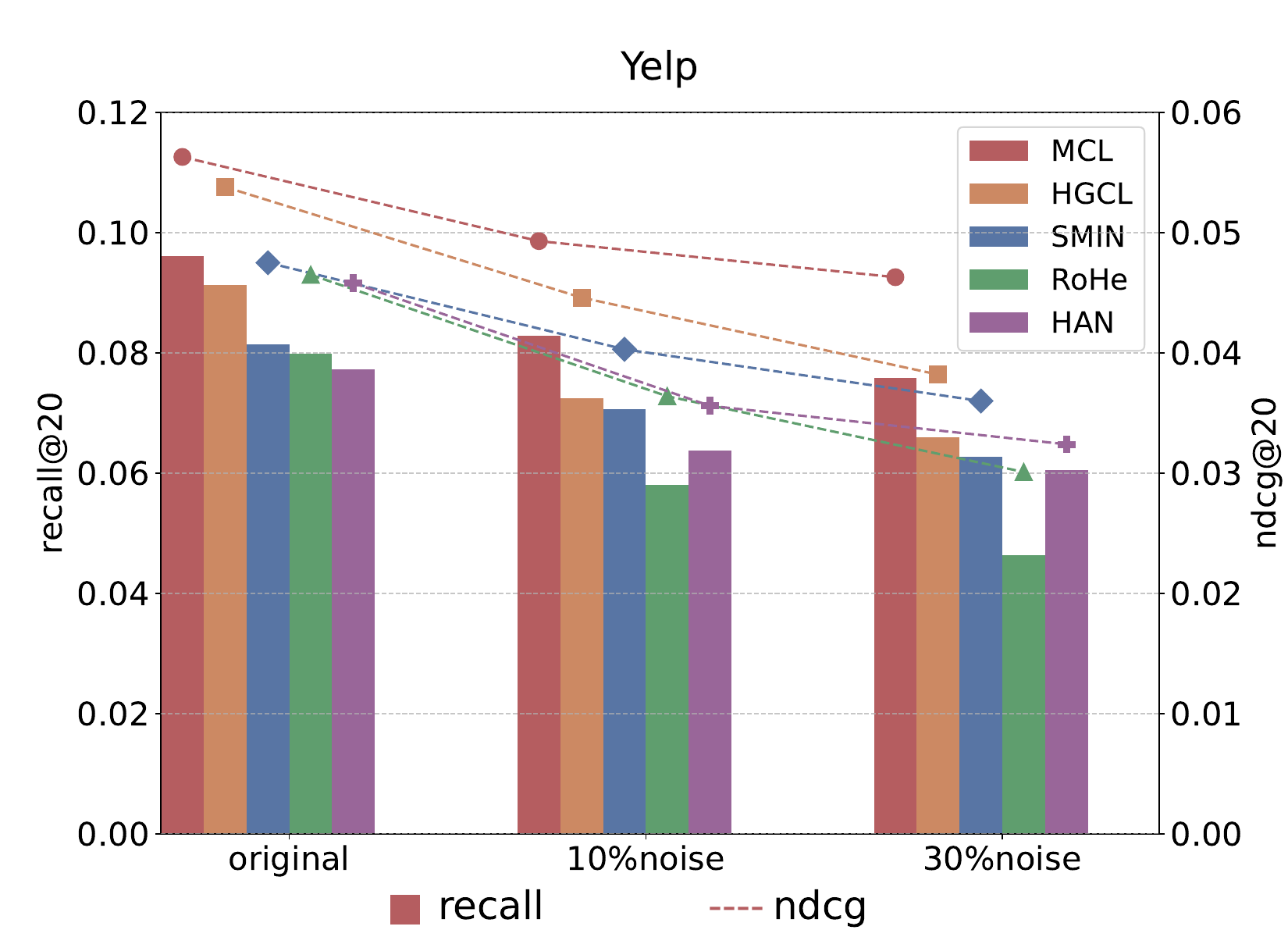}
    \end{minipage}
    \caption{Performance under increasing noise interaction. We introduce 30\% random noise and 10\% noise for comparison to further explore the ability of MCL against noise. On all three datasets, the histogram represents recall@20 and the line graphs represent NDCG@20. The combined analysis of the declining trends in both metrics demonstrates the model's robustness.}
    \label{fig4:random noise}
\end{figure*}

\begin{figure*}[htp]
    \centering
    \begin{minipage}[b]{0.33\linewidth}
        \centering
        \includegraphics[width=\linewidth]{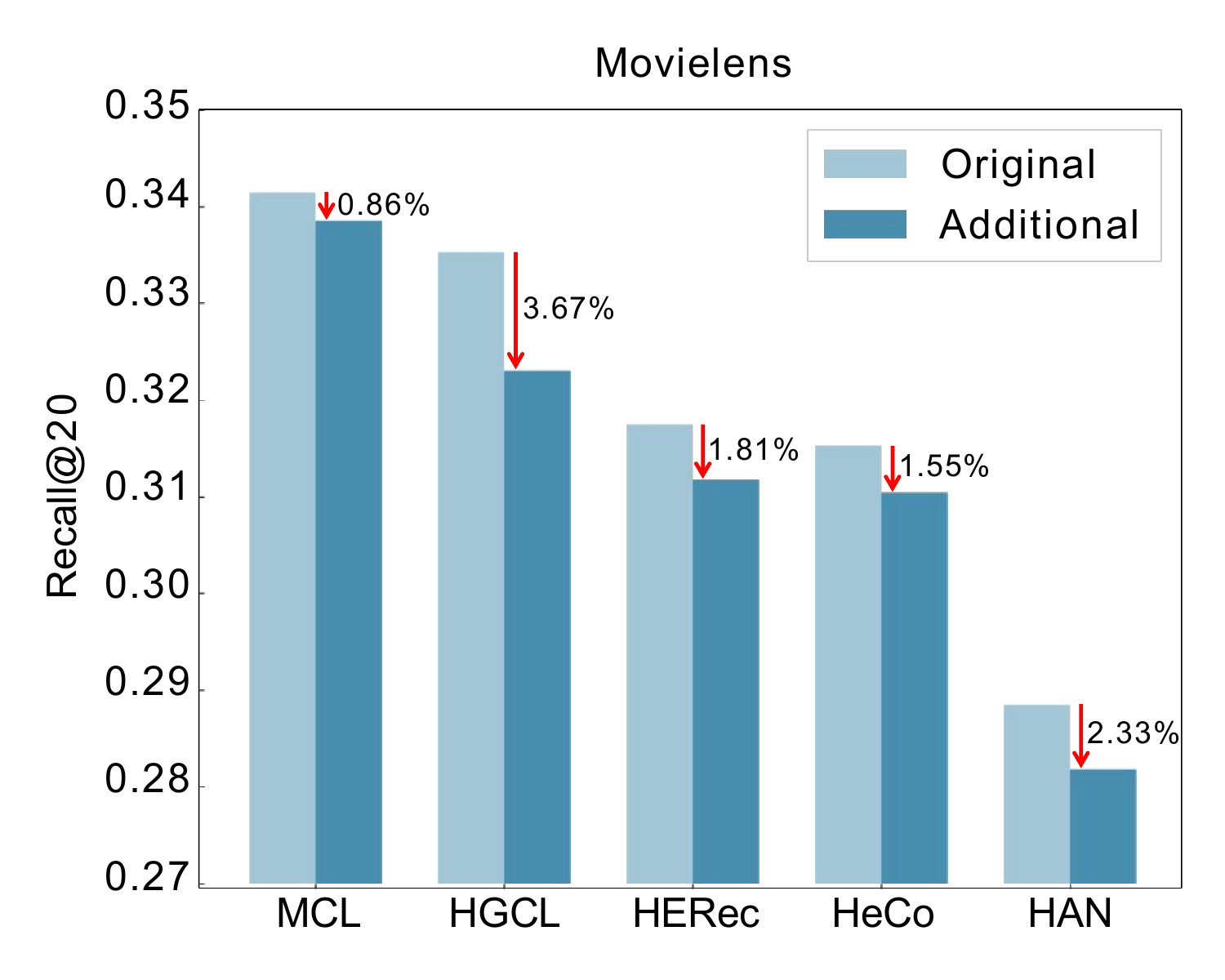}
    \end{minipage}\hfill
    \begin{minipage}[b]{0.33\linewidth}
        \centering
        \includegraphics[width=\linewidth]{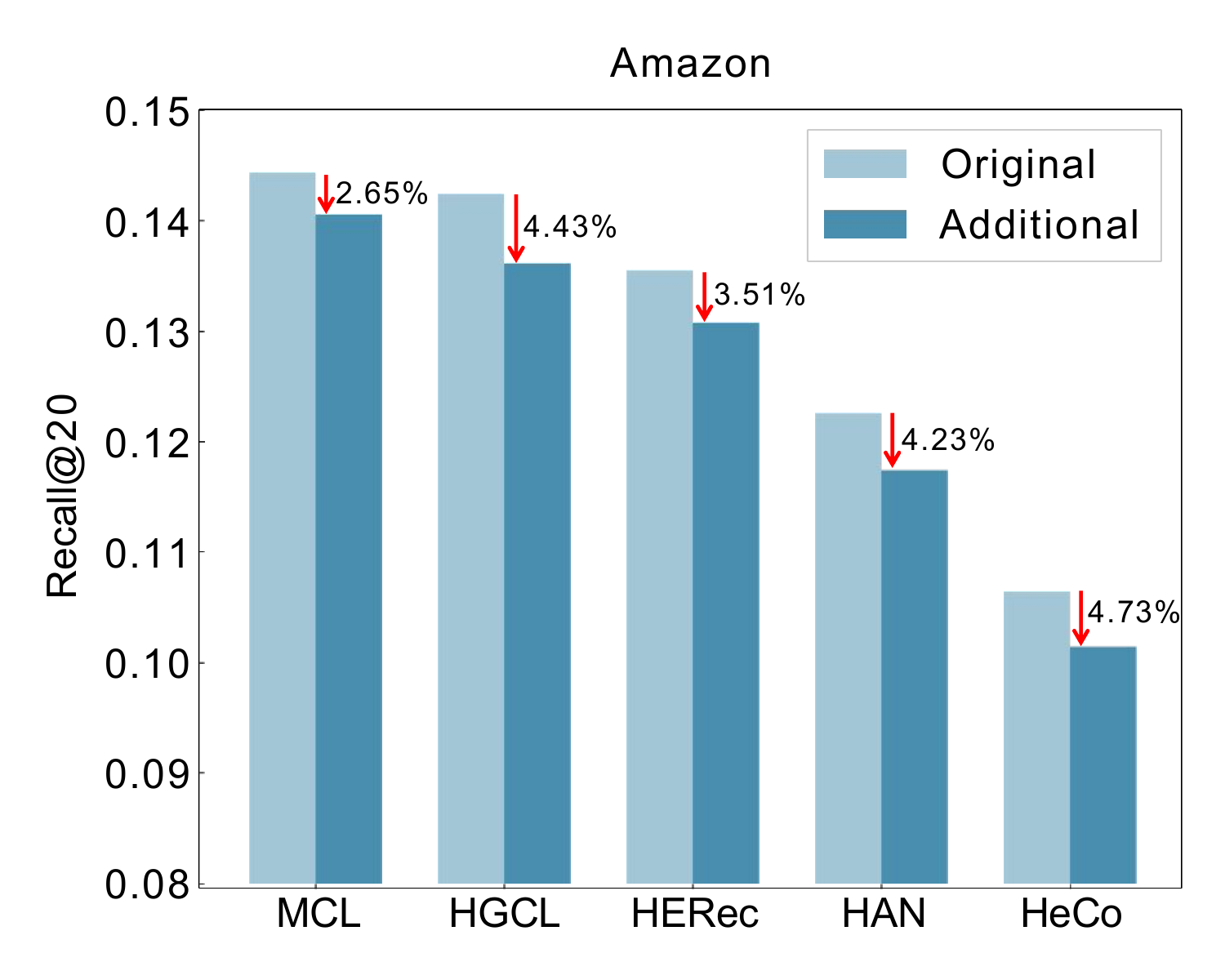}
    \end{minipage}
    \begin{minipage}[b]{0.33\linewidth}
        \centering
        \includegraphics[width=\linewidth]{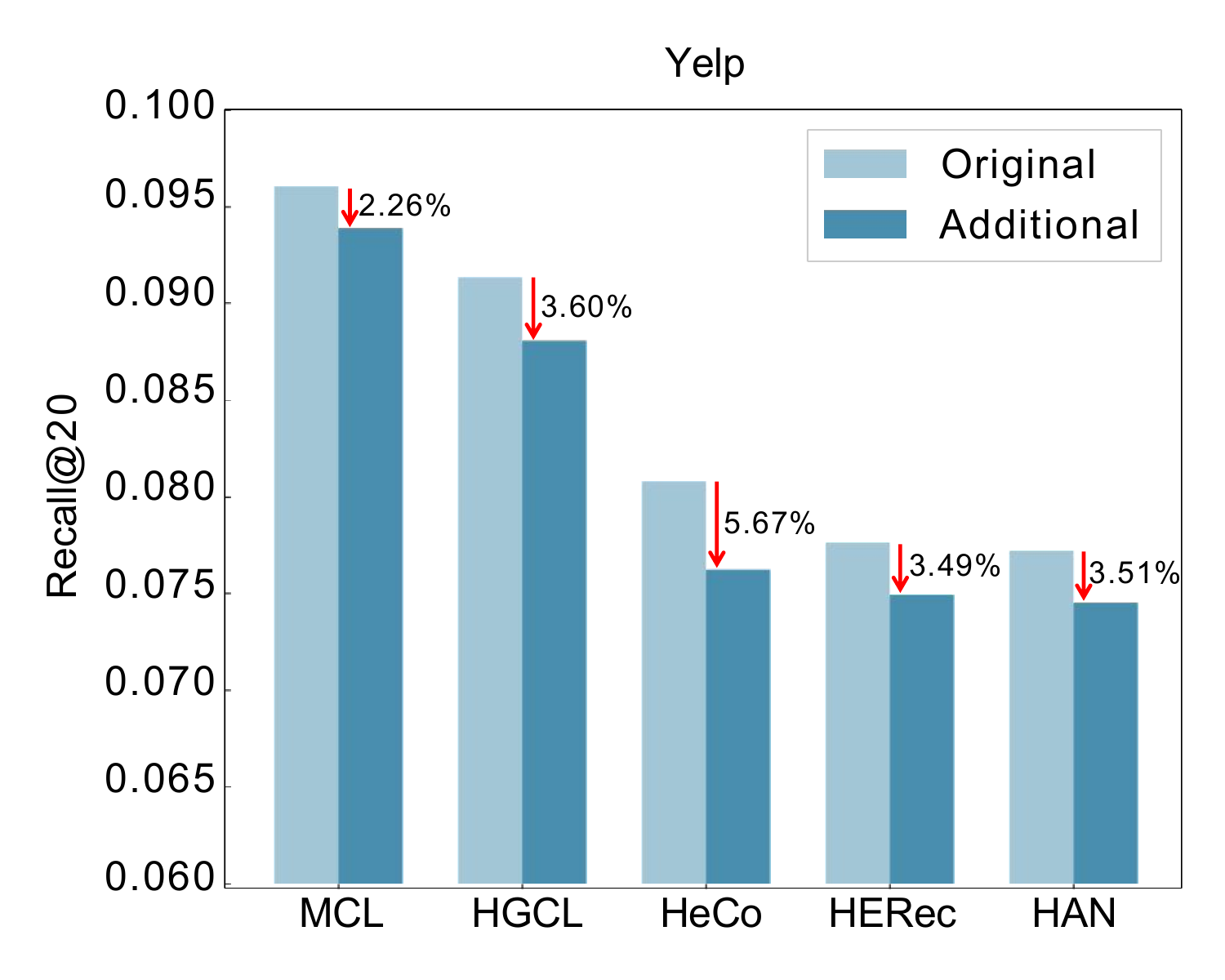}
    \end{minipage}
    \caption{We evaluated the performance after incorporating the meta-paths presented in Table \ref{table5:metapaths}. The light-colored histogram represent the results of the original meta-paths, while the darker colors indicate the results after additional meta-paths. Notably, MCL exhibits only a minor performance decline across all datasets, underscoring the effective noise mitigation of our model in the presence of multi meta-paths.}
    \label{fig5:metapath_noise}
\end{figure*}

\subsection{Robustness of the MCL (RQ2)}
\noindent In this section, we first test the robustness of the model against noisy interactions. Second, we discuss the challenges associated with selecting meaningful meta-paths for recommendation models and explore the impact of including all possible meta-paths. Finally, we conduct an analysis of robustness.
\subsubsection{Robustness to Noisy Interactions}
\noindent To further evaluate the robustness of our model, we added noise interactions to the user-item interactions in the dataset without introducing any additional users. Specifically, we randomly selected 10\% and 30\% of the user-item interactions in the training set for each dataset. For each selected user u in the $\langle u, i\rangle$ interactions, we chose an item j with which user u had never interacted to create a noise edge $\langle u, j\rangle$. This suggests that due to the improved incorporation of masked random propagation, MCL helps the two-level attention model to identify informative graph structures and diminishes its reliance on certain edges. All in all, MCL provides a different perspective to denoise for recommendations. Fig. \ref{fig4:random noise} presents the experimental results on three datasets.

\begin{itemize}[leftmargin=*]
\item As is evident, adding random noise edges to the data leads to a degradation in the performance of both MCL and the baseline model. However, the performance degradation trend of MCL is lower than that of the baseline model. Moreover, as the noise ratio increases, the contrast in the extent of degradation becomes more apparent, as demonstrated by the histograms and line plots in Fig. \ref{fig4:random noise}.
\item We can observe that MCL demonstrates greater robustness on the Amazon and Yelp datasets. The possible reason for this may be that these two datasets are much sparser than movielens-100k. Accordingly, the additional noise has a more pronounced impact on the graph structure, but MCL exhibits enhanced performance. This observation is significant, as sparse data frequently arises in real-world recommendation scenarios, highlighting both the robustness and the potential of MCL.
\end{itemize}

\begin{table}[t]
\caption{Meta-paths selection in robust experiments}
\begin{adjustbox}{width=0.48\textwidth}
\begin{NiceTabular}{|c|c|c|c|}
\toprule $\begin{array}{c}\text { Dataset } \\
\text { (Sparsity) }\end{array}$ & $\begin{array}{c}\text { Relations } \\
\text { (A-B) }\end{array}$ & $\begin{array}{c}\text { Original } \\
\text { meta-paths }\end{array}$ & $\begin{array}{c}\text { Add } \\
\text { meta-paths }\end{array}$ \\
\midrule $\begin{array}{c}\text { Movielens } \\
(93.69 \%)\end{array}$ & $\begin{array}{c}
\text { User-Movie } \\
\text { User-Age } \\
\text { User-Occupation } \\
\text { Movie-Genre } \\ \end{array}$ & $\begin{array}{c}UMU \\ MUM \\ MGM\end{array}$
& $\begin{array}{c}UMGMU \\ UAU \\ UOU \\ MUAUM\end{array}$ \\
\midrule $\begin{array}{c}\text { Amazon } \\
(98.84 \%)\end{array}$ & $\begin{array}{c}
\text { User-Item } \\
\text { Item-View } \\
\text { Item-Category } \\
\text { Item-Brand  } \\ \end{array}$ & $\begin{array}{c}UIU \\ IUI \\ IBI \\ ICI\end{array}$
& $\begin{array}{c}UIBIU \\ UICIU \\ IVI \end{array}$ \\
\midrule $\begin{array}{c}\text { Yelp } \\
(99.91 \%)\end{array}$ & $\begin{array}{c}
\text { User-Business } \\
\text { User-Compliment } \\
\text { Business-City } \\
\text { Business-Category  } \\ \end{array}$ & $\begin{array}{c}UBU \\ BUB \\ BCiC \\ BCaB\end{array}$
& $\begin{array}{c}UCU \\ UBCiBU \\ UBCaBU \end{array}$ \\
\bottomrule
\end{NiceTabular}
\end{adjustbox}
\label{table5:metapaths}
\vspace{-0.5em}
\end{table}

\subsubsection{Robustness to Redundant Meta-paths}
\noindent In many cases, it is difficult to choose meta-paths appropriately. This is because selecting meta-paths reasonably requires experts to understand the semantics introduced by each meta-path and to continuously test the performance of different meta-path combinations, which is a very time-consuming and labor-intensive task. In the selected heterogeneous graph datasets (Table \ref{table5:metapaths}), Relation denotes the type of edge. We have two sets of meta-paths: original meta-paths used in previous experiments and additional meta-paths planned for robust experiments. The original meta-paths tend to introduce useful information and lead to better performance. However, additional meta-paths are difficult to utilize and lead to performance degradation. Why did we opt not to include all the meta-paths in our previous experiments? There are two key reasons. First, after considering the experimental results of all the baselines, the optimal results are mostly achieved under these chosen meta-paths. Second, multiple types of edges give us more meta-path combinations and potential noise. Our chosen models are representative and advanced, encompassing contrastive learning, hierarchical attention, and pre-training.
Considering the findings presented, we have already mentioned that high-order meta-paths, such as `UMGMU', tend to exhibit higher densities than low-order meta-paths such as `UMU'. Moreover, too few nodes in the middle of a meta-path can cause the same problem. For example, in `UAU', Age is only divided into nine classes in the dataset. There are only nine nodes, so many users inevitably interact through `UAU'. Therefore, the process of fusing multiple meta-paths also results in more noise. To enhance the robustness evaluation of our MCL, after adding all meta-paths in the three datasets and comparing them with the current classical model, we can draw the following findings based on the results in Fig \ref{fig5:metapath_noise}.

\begin{itemize}[leftmargin=*]
\item Models \cite{han, rohe}, which rely solely on meta-path subgraphs for information aggregation, exhibit a notable performance degradation. Conversely, models leveraging the entire heterogeneous graph information or user-item interaction information exhibit distinct advantages. Adopting a global or user-item direct interaction perspective is more effective in handling dense meta-path subgraphs.

\item Self-supervised contrastive learning methods \cite{smin, hgcl} have demonstrated great potential for denoising tasks by enhancing the discrimination between different node representations. Therefore, data augmentation and pre-training approaches can uncover more supervised signals from the raw graph data, allowing the graph neural network to learn better node representations.

\item Considering all of the above, our MCL utilizes complementary information from two views: one-hop and meta-path. Additionally, we adopt a combination of embedding enhancement and self-supervised contrastive learning to achieve better robustness and address the problem that meta-paths introduce noise.
\end{itemize}

\subsubsection{Robustness Analysis}
\noindent The robustness against noisy interactions and redundant meta-paths mentioned above \cite{ma2024robust} can be attributed to the following two factors:

\begin{itemize}[leftmargin=*]
\item To account for the structural effect, one can consider removing entire feature vectors of selected nodes rather than dropping out individual feature elements. Random mask allows each node to aggregate information from a subset of its (multi-hop) neighbors by disregarding the features of specific nodes. The random mask and propagation strategy reduce the dependence on particular neighbors, thus enhancing the model's robustness. Empirically, our MCL is shown to produce more stochastic data augmentations and obtain better performance compared to dropout.

\item Over-smoothing occurs in graph neural networks when information aggregation over multiple layers causes the loss of discriminative features, leading to indistinguishable node representations. However, the random mask and propagation strategy allow the model to incorporate additional local information, thereby mitigating the risk of over-smoothing when compared to direct usage of a graph attention network. 
\end{itemize}

\begin{figure}[t]
    \centering
    \begin{minipage}[b]{0.32\linewidth}
        \centering
        \includegraphics[width=\linewidth]{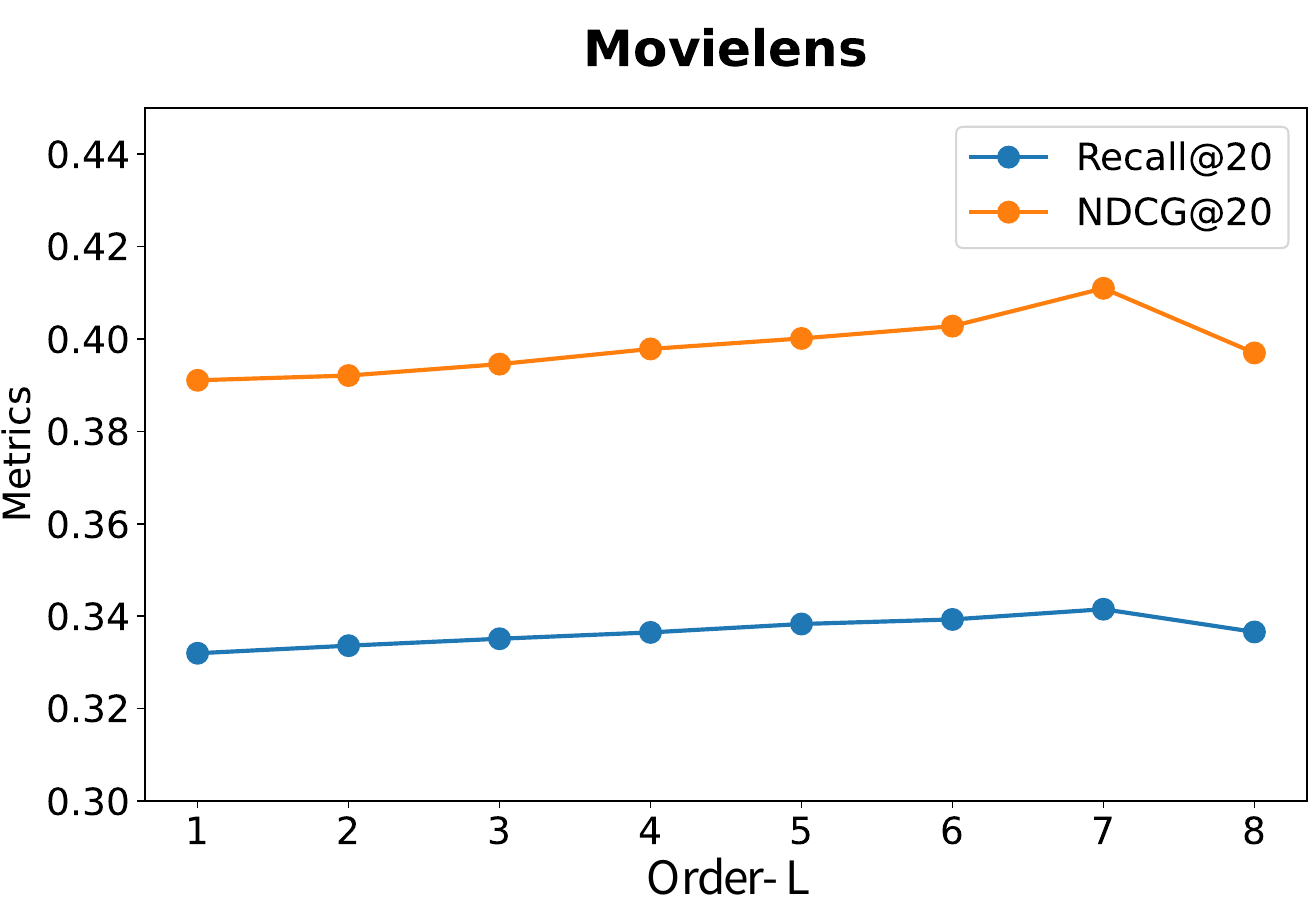}
    \end{minipage}\hfill
    \begin{minipage}[b]{0.32\linewidth}
        \centering
        \includegraphics[width=\linewidth]{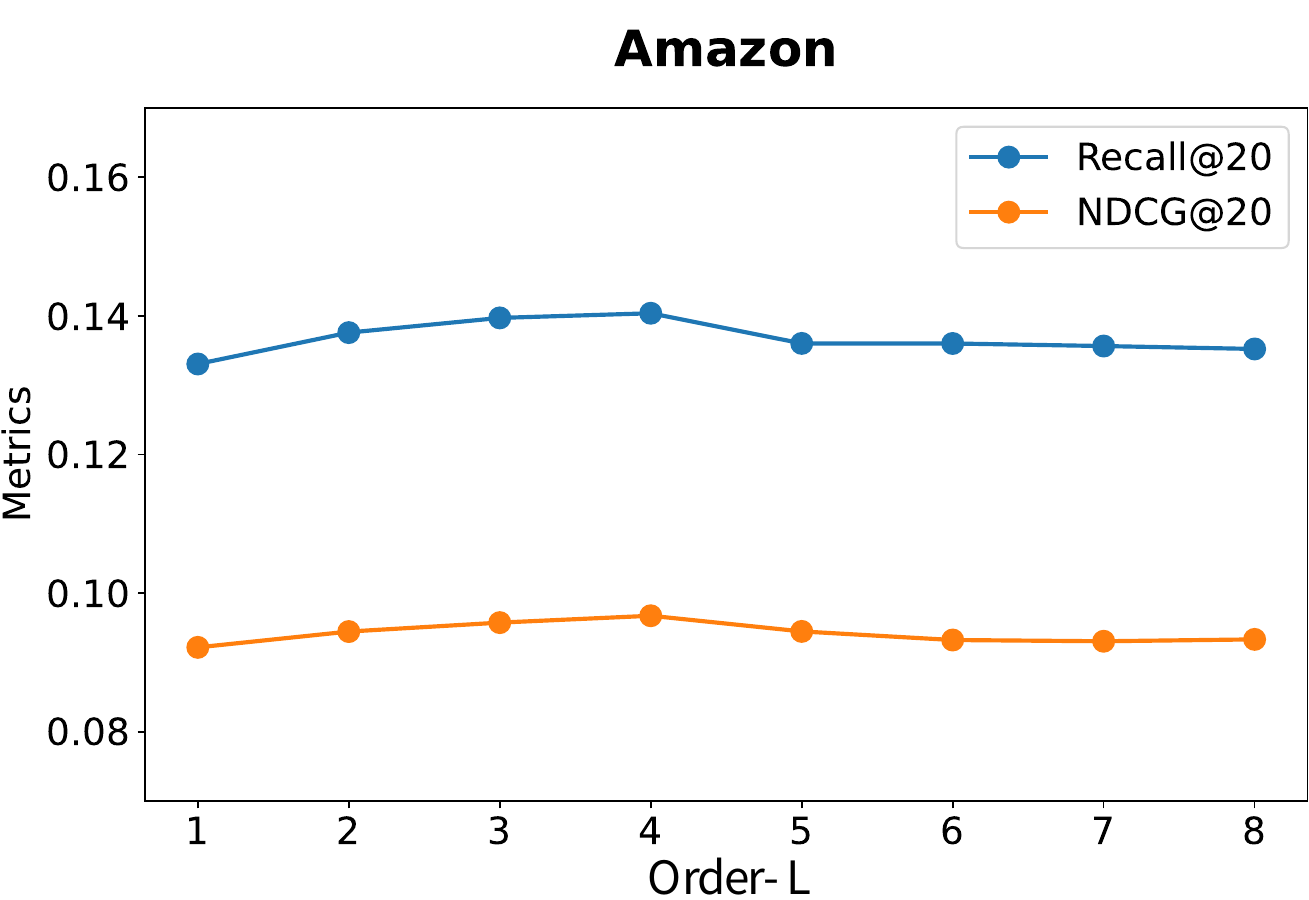}
    \end{minipage}
    \begin{minipage}[b]{0.32\linewidth}
        \centering
        \includegraphics[width=\linewidth]{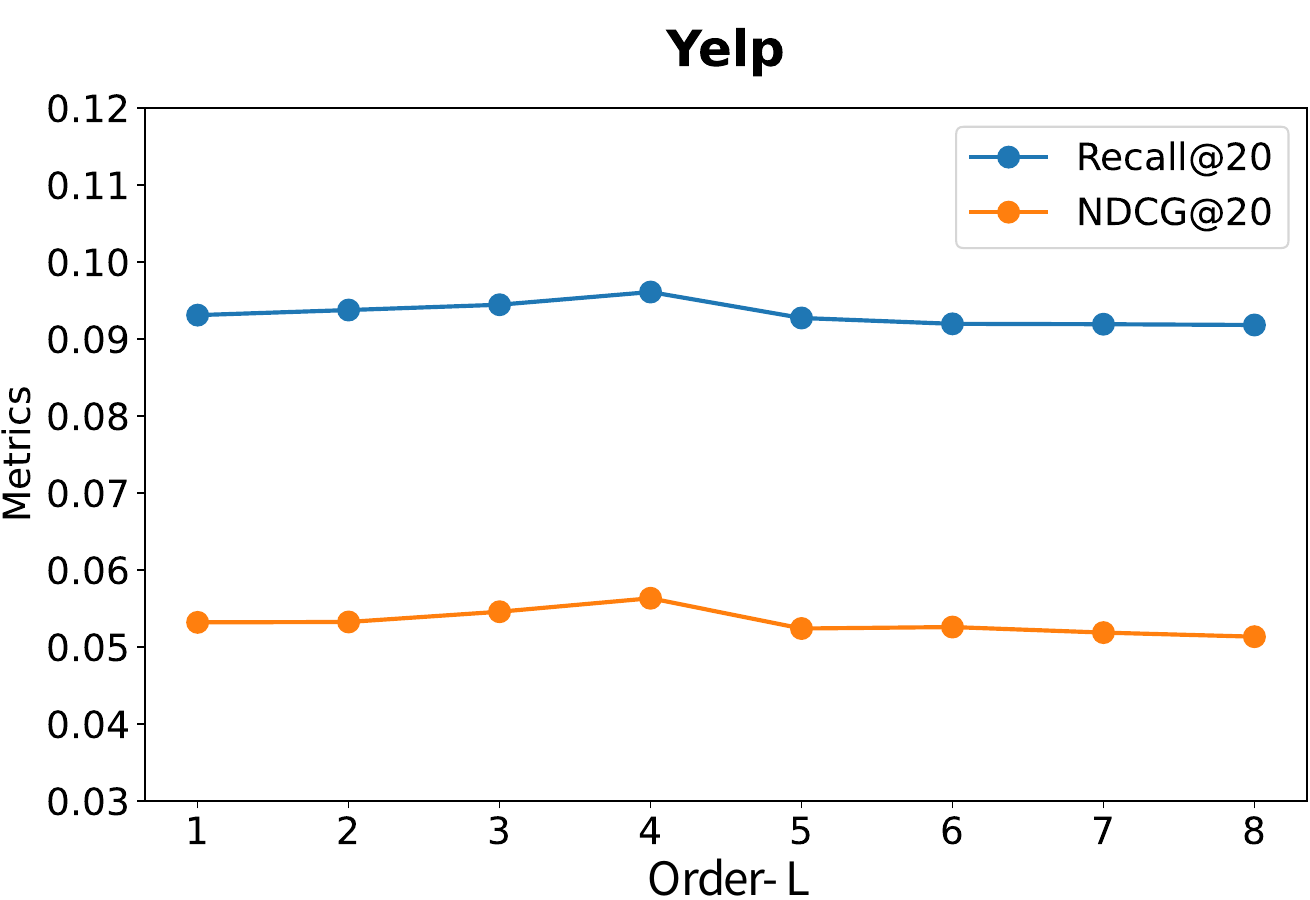}
    \end{minipage}
    \caption{Performance change with respect to the order L of random propagation.}
    \label{fig6:order_K}
\end{figure}

\begin{figure}[t]
    \centering
    \begin{minipage}[b]{0.5\linewidth}
        \centering
        \includegraphics[width=\linewidth]{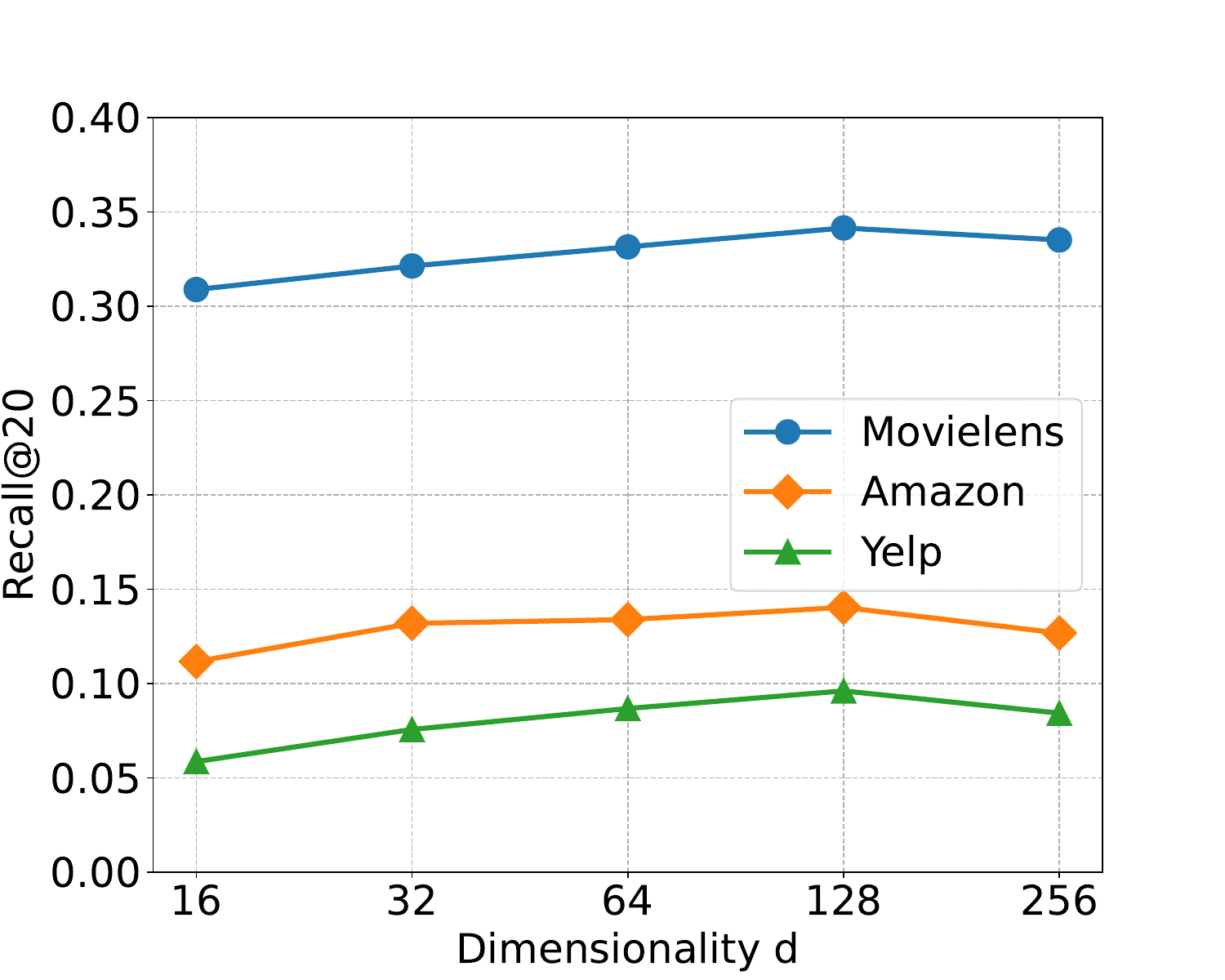}
    \end{minipage}\hfill
    \begin{minipage}[b]{0.5\linewidth}
        \centering
        \includegraphics[width=\linewidth]{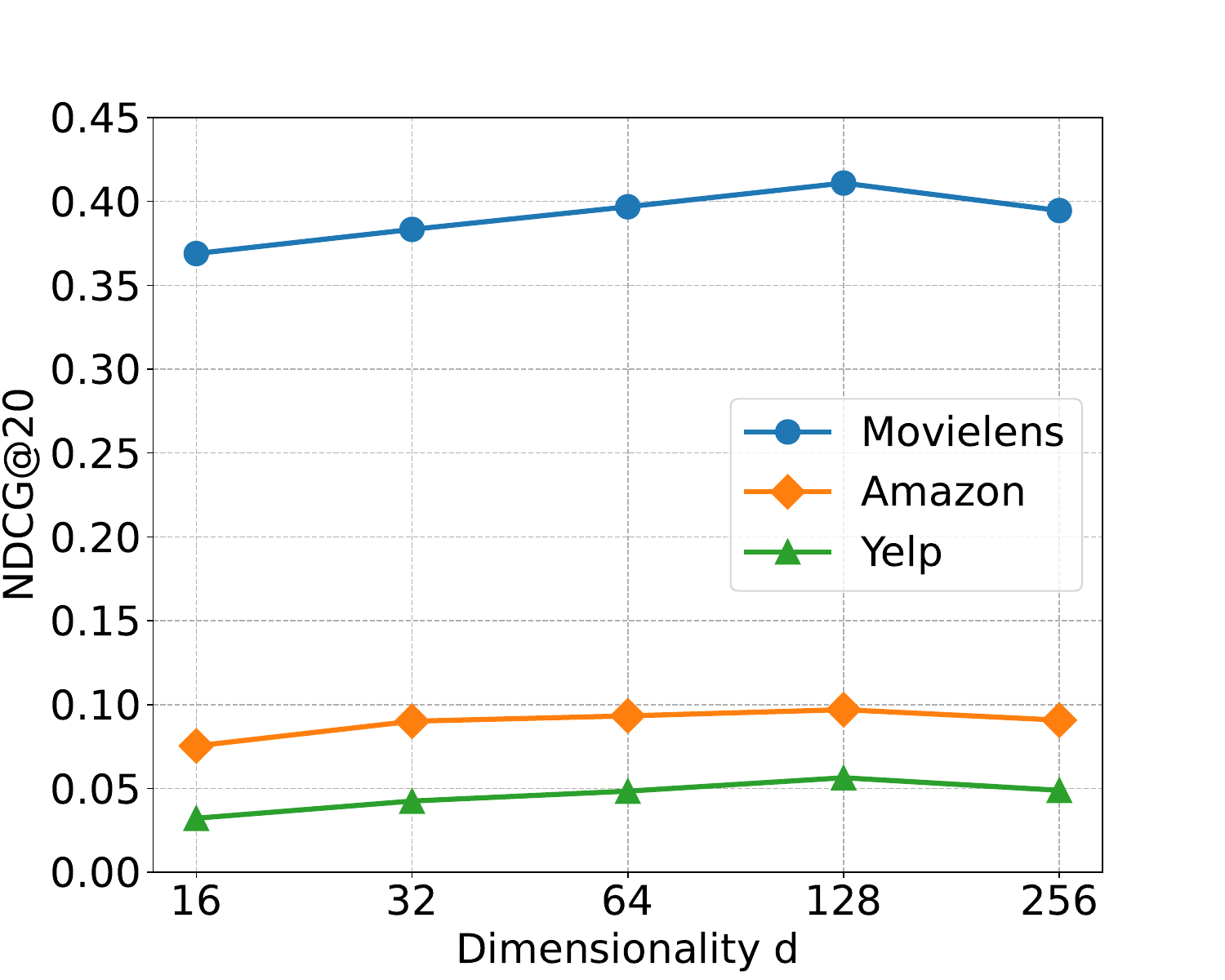}
    \end{minipage}
    \caption{Performance change with respect to the dimension of embeddings on three datasets.}
    \label{fig7:emb}
\end{figure}

\subsection{Ablation Study (RQ3)}
\noindent We conducted ablation experiments to validate the importance of the key components in our model  and provided reasonable explanations or conjectures for the corresponding experimental results.

\begin{itemize}[leftmargin=*]
\item$\boldsymbol{w} / \boldsymbol{o}-\mathbf{mask}$: 
We disabled the  mask and random propagation operation to enhance the embedding. Instead, we directly utilize the randomly initialized embedding to construct the cross-view for the embedding learning process. In other words, we do not include  the part in front of the two-level attentional mechanism shown in Fig. \ref{fig3:framework} (c), but utilize the cross view for contrastive learning as well.

\item $\boldsymbol{w} / \boldsymbol{o}-\mathbf{cl}$: 
We disabled the contrastive learning, an approach that selects enhanced positive samples through efficient and novel sampling. Removing contrastive learning across views to enhance the supervised signals is equivalent to removing part (d) of Fig. \ref{fig3:framework} for the node representation learning of users and items.

\begin{table}[t]
\caption{Ablation study on key components of MCL}
\begin{adjustbox}{width=0.48\textwidth}
\begin{tabular}{ccccccc}
\toprule[1pt]
\midrule Data & \multicolumn{2}{|c|}{ Movielens } & \multicolumn{2}{c|}{ Amazon } & \multicolumn{2}{c}{ Yelp } \\
\midrule Metric & Recall & NDCG & Recall & NDCG & Recall & NDCG \\
\midrule w/o-mask & 0.3290 & 0.3823 & 0.1166 & 0.0806 & 0.0824 & 0.0477 \\
\midrule w/o-cl & 0.3318 & 0.3916 & 0.1350 & 0.0942 & 0.0927 & 0.0546 \\
\midrule w/o-1hop & 0.2848 & 0.3446 & 0.1216 & 0.0825 & 0.0677 & 0.0381 \\
\midrule w/o-meta & 0.3321 & 0.3953 & 0.1305 & 0.0920 & 0.0776 & 0.0425 \\
\midrule MCL & $\mathbf{0.3415}$ & $\mathbf{0.4109}$ & $\mathbf{0.1441}$ & $\mathbf{0.101}$ & $\mathbf{0.0961}$ & $\mathbf{0.0563}$ \\
\midrule
\bottomrule[1pt]
\end{tabular}
\end{adjustbox}
\label{table6:study}
\end{table}

\item $\boldsymbol{w} / \boldsymbol{o}-\mathbf{meta}$: 
In this variant, we exclude the view of learning embeddings from the meta-path subgraphs $\mathcal{G}_{u u}$ and $\mathcal{G}_{i i}$ of users and items using contrastive learning and embedding augmentation. This is equivalent to using only the view of the attention mechanism based on the one-hop neighbor Fig. \ref{fig3:framework} (b) across the HIN.

\item $\boldsymbol{w} / \boldsymbol{o}-\mathbf{1hop}$: 
In this variant, contrary to the ``w/o-meta'' mentioned above, we utilize only views based on meta-path subgraphs to assist in the embedding process, excluding part (b) mentioned in Fig. \ref{fig3:framework}.
\end{itemize}

Table \ref{table6:study} presents the recommendation performance of the MCL model and the compared variants. First, we can see from the table that the performance improvement introduced by the mask component is the most obvious in the three datasets. One possible reason for this phenomenon is the untapped potential within the subgraphs obtained through meta-paths. Effectively utilizing such subgraphs is crucial and demonstrates the effectiveness of our embedding enhancement method. Moreover, contrastive learning also results in performance gains, although they are less prominent than those attained by the mask component. In all cases, MCL outperforms ``w/o-cl'', reflecting that heterogeneous graph contrastive learning is likely to be an effective enhancement for cross-view knowledge transfer. Notably, the key difference between w/o-meta and w/o-mask is that w/o-meta yields inferior results compared to w/o-mask on both the Movielens and Amazon datasets. This observation suggests that the representations learned from the meta-path subgraphs are noisy, thereby validating the problem depicted in `CH1' of the Introduction section.

\begin{figure*}[t]
    \centering
    \begin{minipage}[b]{0.32\linewidth}
        \centering
        \subfloat[\small MCL]{\includegraphics[width=\linewidth]{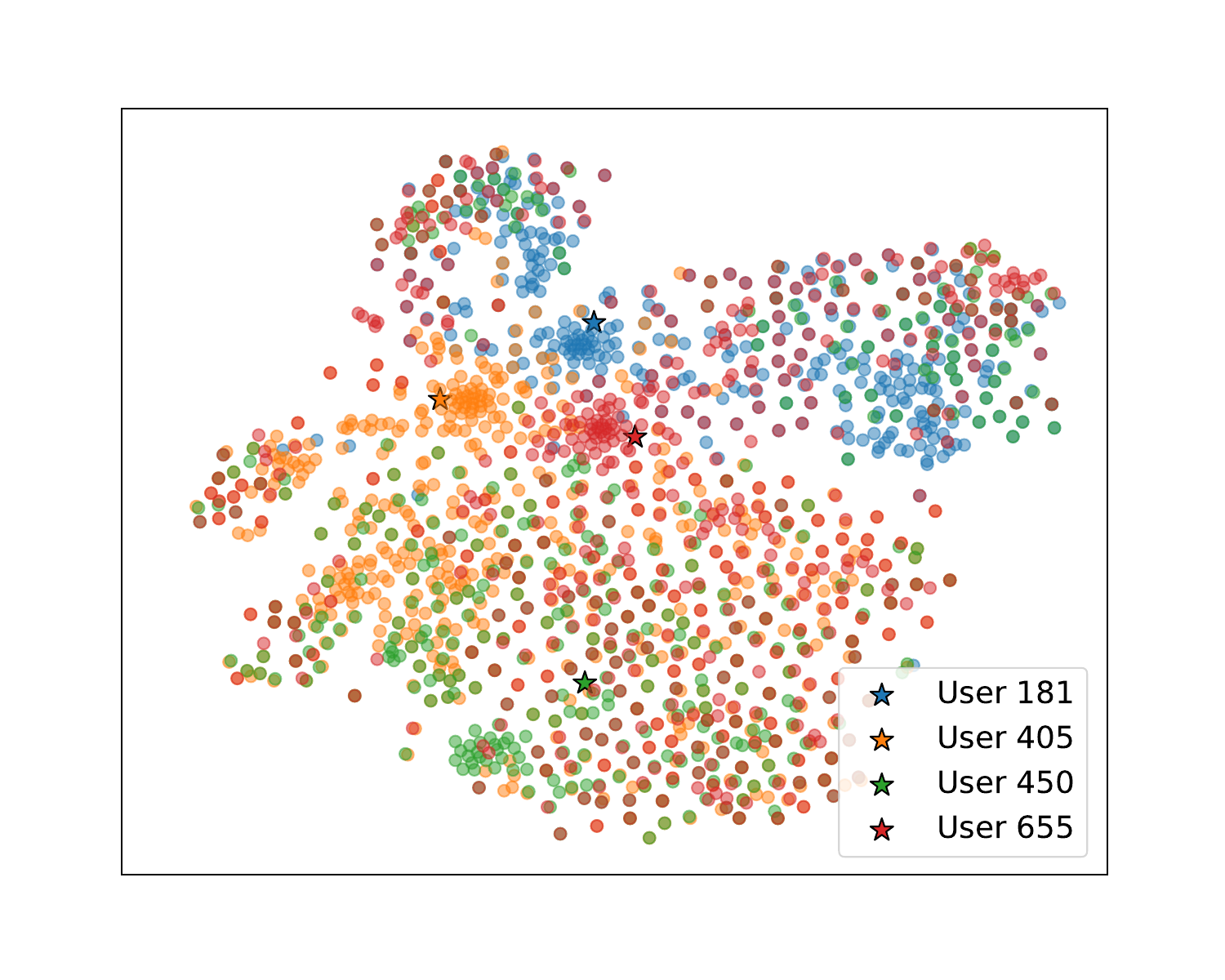}}
    \end{minipage}\hfill
    \begin{minipage}[b]{0.32\linewidth}
        \centering
        \subfloat[\small SMIN]{\includegraphics[width=\linewidth]{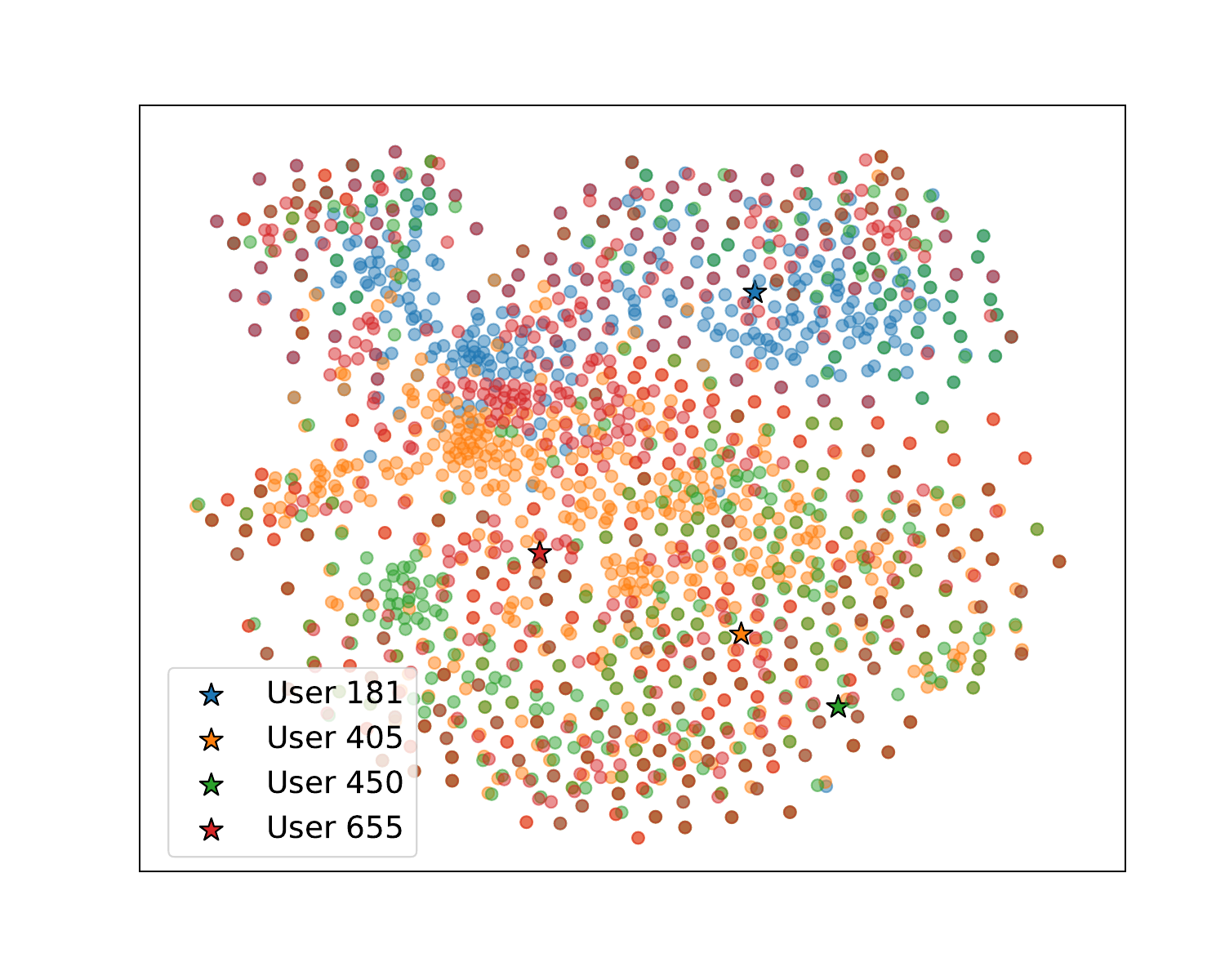}}
    \end{minipage}\hfill
    \begin{minipage}[b]{0.32\linewidth}
        \centering
        \subfloat[\small HAN]{\includegraphics[width=\linewidth]{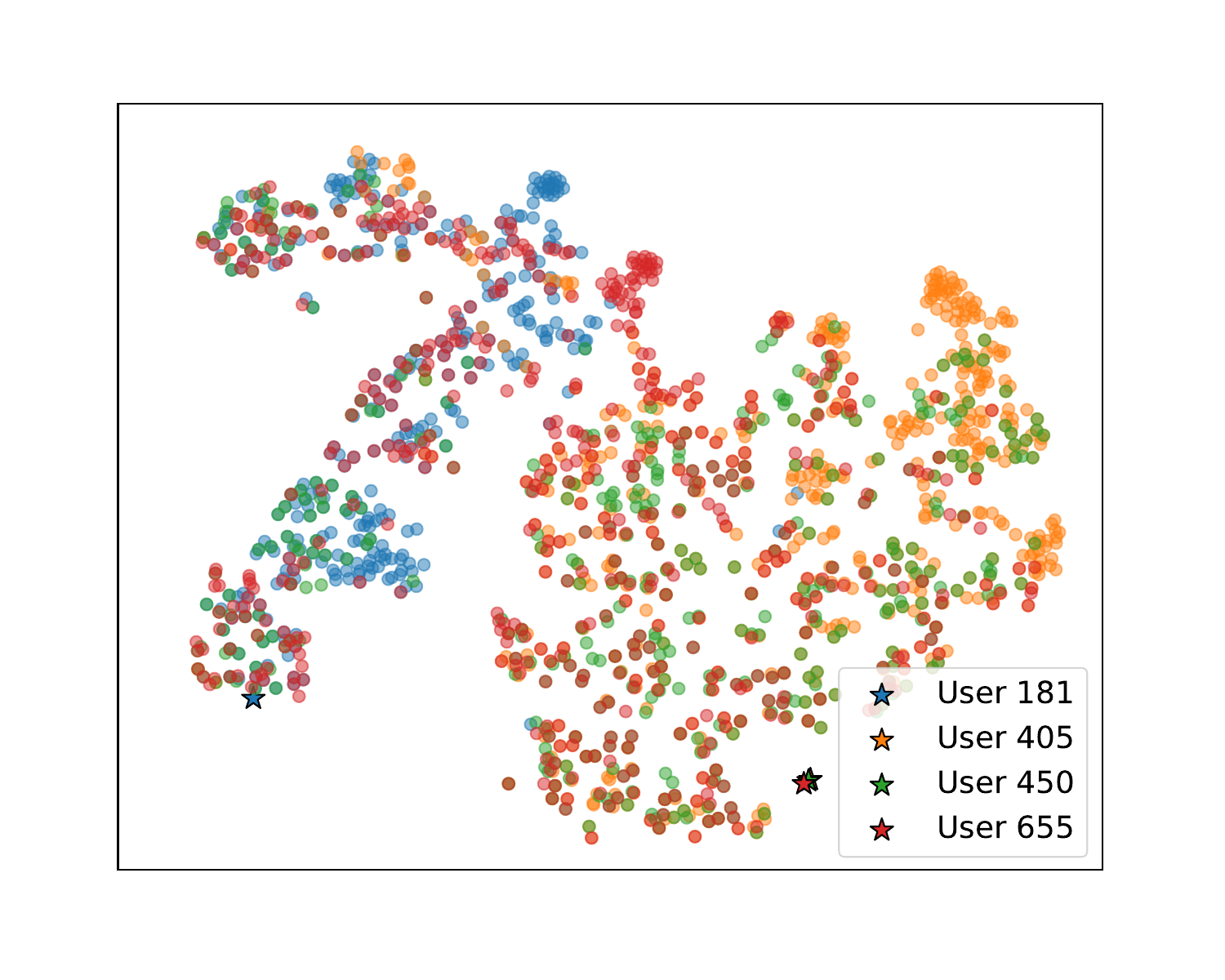}}
    \end{minipage}
    \caption{Embedding visualization of users (stars) and interacted items (circles) based on different encoding methods.}
    \label{fig8:visualization}
    \vspace{-0.1cm} 
\end{figure*}

\subsection{Hyperparameter Analysis}
\noindent We further perform parameter sensitivity analysis from four perspectives: embedding dimension, order L of random propagation, temperature coefficient settings, and learning rate. Selected results are shown in Fig. \ref{fig6:order_K} and \ref{fig7:emb}. Based on the results, we draw the following conclusions.

\begin{itemize}[leftmargin=*]
\item \textbf{Embedding dimension.} We varied the embedding dimension of nodes in the heterogeneous graph $d$ from 16 to 256. We observed that the model's performance reaches its peak when the embedding dimension is set to 128, but further increase leads to a decline in performance. Therefore, the embedding dimension can be increased appropriately to improve the performance of the recommendation model. However, there is a limit to this increase in dimension due to the overfitting of graph neural networks.

\item \textbf{Order L of random propagation.} In our process of embedding perturbations, we can set the number of random propagation steps, as shown in Fig. \ref{fig6:order_K}. Increasing the number of orders enhances the feature enhancement achieved through random propagation. However, many GNNs face the issue of over-smoothing, where the embeddings of different nodes become indistinguishable as the number of propagation steps increases. Based on the decreasing trend observed after reaching L orders, we selected Optimal orders as the optimal number for training in our experiments.

\item \textbf{Contrastive learning temperature coefficient settings.} In our contrastive learning method, there are balancing coefficients $\lambda_{1}$ and temperature coefficients $\tau$. The lambda coefficient is mainly used to balance the weights of the one-hop view and meta-path view for embedded contrastive learning, while the temperature coefficient serves to regulate the degree of attention paid to difficult samples. The temperature coefficient tau is tuned from 0.5 to 0.9 with a step size of 0.05.

\item \textbf{Learning rate.} To investigate the impact of learning rate, we conduct comparative experiments by using different learning rate settings (0.001, 0.005, 0.01, 0.05, 0.1). The optimal learning rate is used for training on different datasets, and other contrast methods are kept under the same optimal learning rate. 
\end{itemize}

\subsection{Visualization of User and Item Representations}
\noindent We use the t-SNE \cite{tsne} algorithm to project the latent representations of users and the items with which they have interacted onto a two-dimensional plane, enabling a clearer visualization of the relationships between them. This two-dimensional mapping helps us better understand user and item interaction patterns and relationships. As shown in Fig. \ref{fig8:visualization}, the embeddings learned by MCL tend to be better (denoted with the same color) than those of HAN and SMIN.
SMIN exhibits a more uniform performance but lacks clustering, while HAN exhibits excessive clustering, making it difficult to distinguish.
This indicates that our approach can effectively preserve user-item interaction relationships effectively. Additionally, we can observe an improved embedding separation phenomenon for distinguishing users interacting with different items. These observations further validate the rationality of our graph-structured main embedding space. 
Mainstream thinking \cite{uniformity, simgcl} has moved beyond the idea that ``the more clustering, the better". The strength of contrastive learning lies in the uniformity of the distribution, and a balance must be struck between clustering and distributional uniformity. Moreover, the contrastive loss is designed to push positive sample nodes away from all negative sample nodes, this also leads to embedding representations in the feature space being closer to a uniform distribution. In recommendation, this strategy gives less popular items more opportunities to be discovered by users, effectively mitigating the popularity bias issue.

\section{Related Work}
\noindent This section discusses three key areas. First, we explore Heterogeneous Information Networks (HINs) for enhanced recommender systems, as they incorporate diverse auxiliary data and enhancing interpretability. Second, contrastive Learning methods are examined for enriching user representation learning in recommender systems. Finally, we discuss Heterogeneous Graph Learning as a crucial approach to effectively process heterogeneous graph data, offering improved node representation and task inference.

\subsection{HIN-based recommender systems}
Recommender systems can be seen as a task of link prediction on information networks. In conventional approaches, some recommender systems employ a bipartite graph model; however, this strategy makes it challenging to effectively leverage diverse auxiliary information. While some researchers have proposed strategies to integrate specific types of auxiliary information, these usually lack generalizability. Recently, recommender systems based on Heterogeneous Information Networks (HINs) \cite{hin} have effectively addressed the unified modeling challenge of integrating diverse auxiliary information and user interaction behaviors. This approach not only excels in mitigating data sparsity and cold-start challenges within recommender systems but also markedly enhances their interpretability. Graph Neural Networks (GNNs) exhibit powerful information propagation and aggregation capabilities. Meta-path-based methods (HERec \cite{HERec}, HAN \cite{han}, and HGCL \cite{hgcl}) initially aggregate adjacent features with shared semantics before integrating divergent semantic elements. Meta-path-free methods (e.g., HGT \cite{hgt}) resemble GNNs in aggregating messages from local one-hop neighborhoods without consideration of node types, yet incorporate supplementary modules (e.g., attention mechanisms) for embedding semantic details (e.g., node and edge types) within the propagated messages. To enhance the learning of collaborative relationships between users with social influence, some studies have developed social relationship encoders employing graph neural networks; examples include GraphRec \cite{graphrec}and RGCN \cite{rgcn}. Furthermore, several multimedia recommender systems leveraging graph neural networks (e.g., GRCN \cite{grcn}) have emerged, which integrate diverse modalities of information into the recommendation process.

\subsection{Contrastive Learning for Recommendation}
Recently, researchers have begun to focus extensively on contrastive self-supervised learning, because the generated self-supervised signals can be used to enrich user representation learning. Research on self-supervised learning can be broadly divided into two main branches: generative models \cite{bert, distributed} and contrastive models \cite{sgl}. Contrastive models augment the model by comparing Noise Contrast Estimation (NCE) object and alignment between contrastive views. We now roughly summarize the currently dominant contrastive learning methods. 
These include improvements to graph data enhancement methods, 
multi-view-based contrastive learning methods, contrastive tasks using node relationships (using the relationship between nodes and neighboring nodes as the sample selection criterion, and the relationship between the output representations of GNN nodes in different layers, hypergraphs can be considered), and other perspectives of contrastive learning tasks. 
For example, many studies have addressed the problem of data sparsity in recommenders by proposing various enhancement schemes for contrastive embedding, such as SGL \cite{sgl}, and SimGCL \cite{simgcl}. Among these, SGL utilizes random node/edge dropout to generate views for contrastive learning. SimGCL generates contrast views by adding uniform noise to the embedding space. 
Furthermore, contrastive learning has found applications in various recommendation scenarios, encompassing multi-behavior recommendation \cite{multi_be}, and multi-interest recommendation \cite{multi_in}.

\subsection{Heterogeneous Graph Learning}
Heterogeneous graphs contain rich semantics in addition to structural information, represented by a wide variety of nodes and edges. Representation learning \cite{hrl, wang2023user, liu2021self} for heterogeneous graphs aims to solve the problems of node representation learning and task inference in heterogeneous graph data. In recent years, heterogeneous graph neural networks have been widely explored and made significant progress in this task. For example, to model node relationships more accurately, the concept of the meta-path \cite{meta-path, chen2022meta} has been introduced in some studies. HAN \cite{han} captures inter-node relationships adaptively through the attention mechanism. On the other hand, inspired by the converter framework, HGT \cite{hgt} proposes an innovative graph converter network, which employs the self-attention mechanism to compute the propagation weights between nodes, thereby realizing the effective transfer of heterogeneous information. HGSL \cite{hgsl} aims to jointly learn the structure of heterogeneous graphs and the parameters of graph neural networks. This approach considers the heterogeneous relationships and combines feature similarity, feature propagation, and semantic information within the generated subgraphs to optimize the overall heterogeneous graph structure. HeCo \cite{heco} introduces the cross-view contrastive mechanism, combining the network architecture and meta-path view; as a result, it can effectively learn the node embedding, capture the local and higher-order structure, and enhance the performance of heterogeneous information network analysis. In conclusion, Heterogeneous Graph Learning has made significant research progress as an effective method for processing heterogeneous graph data. Building upon this cutting-edge research direction, the present paper delves into applying Heterogeneous Graph Contrastive Learning to recommender systems, addressing an underexplored task area.

\section{Conclusion}
\noindent In this paper, we focus on investigating the presence of heterogeneity in graph-based recommender systems. We present a novel model called  Mask Contrastive Learning (MCL). This model incorporates contrastive learning techniques by leveraging embedding masks and random propagation to effectively capture user-item representations from two distinct structural perspectives: local and higher-order views. Through extensive experiments on three real-world datasets, we demonstrate that our model outperforms existing approaches in terms of performance and robustness.



\section*{ACKNOWLEDGEMENTS}
This work is supported by the National Natural Science Foundation of China (No. 62272001 and No. 62206002), Anhui Provincial Natural Science Foundation (2208085QF195), and Hefei Key Common Technology Project (GJ2022GX15).

\nocite{*}
\bibliographystyle{IEEEtran}
\bibliography{ref}

\end{document}